\documentclass{article}
\usepackage[utf8]{inputenc}
\usepackage{xcolor}
\usepackage{geometry}
\usepackage{amssymb,amsmath}
\usepackage{graphicx}
\usepackage{bm}
\usepackage{pgf,tikz}
\usetikzlibrary{positioning}
\usepackage{amsthm}
\usepackage[ruled,vlined,linesnumbered]{algorithm2e}

\usepackage{booktabs}
\usepackage{longtable}
\usepackage{array}
\usepackage{multirow}
\usepackage{wrapfig}
\usepackage{float}
\usepackage{colortbl}
\usepackage{pdflscape}
\usepackage{tabu}
\usepackage{threeparttable}
\usepackage{threeparttablex}
\usepackage{ifxetex,ifluatex}
\usepackage[section]{placeins}
\usepackage{adjustbox}
\usepackage{mathtools}
\usepackage{url} 

\usepackage{natbib}
\setcitestyle{authoryear,open={(},close={)}} 

\newcommand{\RR}{\mathbb{R}}

\newcommand{\PP}{\mathbb{P}}

\newcommand{\indep}{\perp \!\!\! \perp}

\newtheorem{theorem}{Theorem}

\allowdisplaybreaks

\usepackage{titlesec}
\titleformat{\paragraph}
{\normalfont\normalsize\bfseries}{\theparagraph}{1em}{}
\titlespacing*{\paragraph}
{0pt}{3.25ex plus 1ex minus .2ex}{1.5ex plus .2ex}

\usepackage{authblk}

\title{Targeted Learning on a Variable Importance Measure for Heterogeneous Treatment Effects}
\date{March 2026}

\author[1]{Haodong Li}

\author[1]{Alan E Hubbard}

\author[2]{Oliver J Hines}

\author[3]{Andrea M Stor{\aa}s}

\author[3]{Kajsa Kvist}

\author[1]{Mark J van der Laan}

\affil[1]{University of California Berkeley, United States}
\affil[2]{Columbia University, United States}
\affil[3]{Novo Nordisk, Denmark}

\begin{document}

\maketitle

\begin{abstract}
   Quantifying the heterogeneity of treatment effect is important for understanding how a commercial product or medical treatment affects different population subgroups. Beyond the main effect parameters like the average treatment effect, the analysis of treatment effect heterogeneity further reveals details on the importance of different covariates and how they lead to different treatment impacts. While much of treatment effect heterogeneity analysis focuses on the conditional average treatment effect, an alternative parameter that captures treatment effect heterogeneity is the variance of treatment effect across different covariate groups. One can also derive variable importance parameters that measure (and rank) how much of treatment effect heterogeneity is explained by a targeted subset of covariates. In this article, we propose a new targeted maximum likelihood estimator (TMLE) for a treatment effect variable importance measure, in the form of the difference of the variances of conditional average treatment effect. This TMLE is a pure plug-in estimator that consists of two steps: 1) the initial estimation of relevant components to plug in and 2) an iterative updating step to optimize the bias-variance tradeoff. Simulation results show that the proposed estimator has competitive performance in terms of lower bias and better confidence interval coverage compared to a simple substitution estimator and an estimating equation estimator. We apply these methods to data from a randomized clinical trial comparing nucleoside monotherapy with combination therapy in HIV-infected adults. We find that the estimating equation estimator and the proposed TMLE suggest similar rankings of variable importance. The application of this method also demonstrates the advantage of a plug-in estimator, which always respects the global constraints on the data distribution and that the estimand is a particular function of the distribution. \\
\end{abstract}

\section{Introduction}
The growth of methods for precision health, as well as targeted advertising, have highlighted the importance of more refined information than average population treatment effects~\citep{MurphyODTR2003a, RothwellSarcJ2005, RosenkranzESACJ2020, SennITNDO2001}. As a result, understanding the heterogeneity of treatment effects has been widely discussed in the statistical and econometric literature~\citep{CuiEHTEA2023, DingDTEVJ2019, HeckmanMMOP1997, KunzelMEHTM2019, WillkeCTEHD2012, AtheyRPHCJ2016, WagerEIHTJ2018}. Estimation of heterogeneous treatment effects can help clinical professionals make decisions tailored to each patient's characteristics, or helps commercial companies to recommend personalised advertisements to their potential customers. To formalize our discussion of treatment heterogeneity, we use the potential outcome causal framework~\citep{Splawa-NeymanAPTA1990, RubinECETO1974}. The average treatment effect (ATE) is a fundamental parameter in causal inference that helps researchers evaluate the main effect of treatment on the whole population. One simple approach to understanding treatment effect heterogeneity is to stratify the population by one or more discrete covariates and estimate the ATE in each of the resulting subgroups. In general, when covariates are potentially continuous, there are other methods to estimate the conditional average treatment effects (CATE)~\citep{AbrevayaECATO2015, AtheyRPHCJ2016, WagerEIHTJ2018}. \\

\noindent Existing testing and estimating approaches for CATE include nearest-neighbor matching, kernel methods, series estimation, forest-based methods, and the adaptive debiased machine learning method~\citep{CrumpNTTEA2008, LeeNTDT2009, WillkeCTEHD2012, WagerEIHTJ2018, vanderLaanADMLJ2023}. In this analysis, we implemented two general meta-learning methods for CATE estimation: 1) the direct method that estimates the conditional expectation of the outcome by using all of the covariates and the treatment variable and then takes the difference between the estimates that differs from the value of treatment variable~\citep{GreenMHTE2012, KunzelMEHTM2019}; 2) a double robust method that regresses the augmented inverse propensity weighted score on the covariates~\citep{vanderLaanTLMOM2015, LuedtkeSODTM2016a}. We will refer to the first CATE estimation method as S-learner and the second method as the DR-learner, following the convention in~\cite{KunzelMEHTM2019}. In addition to S-learner and DR-learner, alternative CATE metalearners including T-learner, X-learner, R-learner~\citep{WagerEIHTJ2018, KunzelMEHTM2019} could alternatively be employed.\\

\noindent Existing work on how to utilize CATE estimates mainly focus on policy learning and optimal dynamic treatment rules (ODTR)~\citep{MurphyODTR2003a, AtheyPLOD2021, KallusMEPLD2020, vanderLaanTLMOM2015, LuedtkeOITRM2016, MontoyaODTRO2021}. The aim of making personalized treatment decision motivates a different type of intervention scheme called a dynamic treatment rule, as opposed to a static rule which always assigns the same intervention regardless of individual's characteristics. Formally, a dynamic treatment rule can be thought of as a function that maps covariates to the set of treatment values~\citep{BembomPiir2007,MurphyODTR2003a}. In existing proposals for ODTR, the treatment decision is often made based on the values of functions of the CATE estimates with the objective of maximizing the treatment effect~\citep{LuedtkeOITRM2016}. In addition to policy learning and ODTR estimation, some CATE estimation methods also can be used for variable importance measures (VIM) estimation. For example, causal forest, a forest-based CATE estimation method developed from random forest, provides variable importance measures relying on the ``tree architecture'' of causal forest models~\citep{BreimanRFO2001, AtheyGRFA2019, AtheyRPHCJ2016, WagerEIHTJ2018}. These measures are inherently tied to choice of a specific forest-based CATE estimator and have also been criticized since they tend to assign greater importance to continuous variables, or categorical variables with many categories~\citep{StroblBRFVJ2007, GrompingVIARN2009, HinesVIMHA2022}. What have been less available to researchers are reliable measures of the importance of covariates in determining the ODTR, or a variable importance measure with regard to the impact on the CATE funciton. \\

\noindent The VIM addressed in this article is specifically related to a variable's impact of the variance of the treatment effect (VTE), a global measure of treatment effect heterogeneity~\citep{LevyFmteJ2021}. In~\cite{HinesVIMHA2022}, the authors proposed two alternative VIMs along with estimating equation (EE) estimators for heterogeneous treatment effects, which connect to the regression-VIMs and the nonparametric VIM framework discussed in~\cite{Williamson2020, Williamson2023}. The first (unscaled) VIM is in the form of the difference of the variances of CATE. The second (scaled) VIM is defined as the first VIM scaled by VTE. Both VIMs are algorithm-agnostic and have similar interpretation on treatment heterogeneity. They are straightforward to explain to researchers who know traditional goodness-of-fit approaches and, unlike algorithmic VIMs, they enable comparison of heterogeneous treatment effect findings across different CATE learners~\citep{HinesVIMHA2022}. In this article, we propose a new targeted maximum likelihood estimator (TMLE) for the unscaled VIM. TMLE is a two-step procedure where one first obtains an estimate of the data-generating distribution $\PP_0$, or the relevant portion of $\PP_0$. The second stage updates this initial fit such that an optimal bias-variance tradeoff for the parameter of interest is achieved~\citep{LaanTLCI2011}. We evaluate the performance of the new TMLE via simulations. We find that the TMLE have competitive performance in terms of lower bias and better nominal confidence interval coverage compared to the EE estimator and the simple substitution (SS) estimator. \\

\noindent In section 2 below, we introduce the data, notations and the target parameters. In section 3, we discuss the estimators, the asymptotic normality, construction, and implementation of the new TMLE. Section 4 presents the simulation results of all three target parameters and compares the performance metrics of SS, EE and TMLE estimators. Section 5 demonstrates the implementation of the TMLE and EE VIM estimators with data from a randomized clinical trial comparing nucleoside monotherapy with combination therapy in HIV-infected adults~\citep{Hammer1996}. Section 6 concludes the article.

\section{The Estimation Problem}

\subsection{Data and Model}
Throughout the discussion in this article, we consider an observed data structure defined as: $O \equiv (W, A, Y) \sim{\PP_0} \in \mathcal{M}$ with $n$ independent and identically distributed (i.i.d.) observations $O_1,...,O_n$, where $W$ denotes baseline covariates; $A$ denotes the treatment variable (binary); $Y$ denotes the outcome variable; $\mathcal{M}$ denotes the statistical model for the probability distribution of the data, which is nonparametric, beyond possible knowledge of the treatment mechanism (i.e. the conditional distribution of A given W); $\PP_0$ denotes the true data-generating distribution. \\

\noindent Underlying the observed data, we assume a structural causal model (SCM): $W = f_W(U_W)$, $A = f_A(W, U_A)$, $Y = f_Y(W, A, U_Y)$, where $U =(U_W,U_A,U_Y)$ denotes a vector of unmeasured errors following a distribution $\PP_U$; $f =(f_W,f_A,f_Y)$ denotes a vector of functions. $\PP_{O,U}$, the full distribution of $(O,U)$ is then parametrized by $\PP_U$ and $f$. We use $\mathcal{M}^F$ to denote the collection of all possible full distributions $\PP_{O,U}$ described by the SCM, satisfying assumptions on $\PP_U$ and $f$. The statistical model $\mathcal{M}$ is linked to $\mathcal{M}^F$ in that the observed $O = (W, A, Y)$ is generated from a $\PP_{O,U}$ described by the SCM. In other words, a statistical model $\mathcal{M}$ can be augmented with additional causal assumptions providing enriched interpretation~\citep{LaanTLCI2011}.


\subsection{Target Causal Parameters}
A few ingredient terms are needed for defining the target causal parameters: first, by intervening the treatment variable $A$, we can define the counterfactual outcomes $Y^1 = f_Y(W, A=1, U_Y)$ and $Y^0 = f_Y(W, A=0, U_Y)$. Then CATE can be defined as $\tau(W) \equiv E(Y^1|W) - E(Y^0|W)$. In addition, we use $\tau_s(W) \equiv E(\tau(W)|W_{-s})$ to denote the CATE given a subset 
of covariates, where $W_{-s} \subseteq W$ is a generic covariate set excluding those 
indexed by $s$. Now we can define the target causal parameters: $\Psi^F_1(\PP_{O,U}) \equiv \mbox{var}(\tau(W))$, $\Psi^F_2(\PP_{O,U}) \equiv \mbox{var}(\tau(W)) - \mbox{var}(\tau_{s}(W))$, and $\Psi^F_3(\PP_{O,U}) \equiv \Psi^F_2(\PP_{O,U}) / \Psi^F_1(\PP_{O,U})$, corresponding to VTE, the unscaled treatment effect VIM and the scaled treatment effect VIM respectively.

\subsection{Identification Assumptions}
With the following assumptions: consistency ($A=a \Rightarrow Y=Y^a$), randomization ($(Y^0, Y^1)  \indep A~|~W$) and positivity ($0<\PP(A=1|W)<1$ almost surely), CATE $\tau(W)$ can be identified by $\tau_0(W) \equiv E(Y|A=1, W) - E(Y|A=0, W) = \bar{Q}(1, W) - \bar{Q}(0, W)$, and $\tau_s(W)$ can be identified by $\tau_{s,0}(W) \equiv E(\tau_0(W)|W_{-s})$. The target causal parameters can be identified as a mapping from $\mathcal{M}$ to $\RR^2$: $(\Psi^F_1(\PP_{O,U}), \Psi^F_2(\PP_{O,U}), \Psi^F_3(\PP_{O,U})) = (\Psi_1(\PP_0), \Psi_2(\PP_0), \Psi_3(\PP_0))$, where $\Psi_1(\PP_0) \equiv \mbox{var}(\tau_0(W))$, $\Psi_2(\PP_0) \equiv \mbox{var}(\tau_0(W)) - \mbox{var}(\tau_{s,0}(W))$, and $\Psi_3(\PP_0) \equiv \Psi_2(\PP_0) / \Psi_1(\PP_0)$. We refer $(\Psi_1(\PP_0), \Psi_2(\PP_0), \Psi_3(\PP_0))$ as the target statistical parameters (or statistical estimands).

\subsection{Target Statistical Parameters}
 In this section, we will re-parametrize the three target statistical parameters and discuss the influence functions. Given that a TMLE for $\Psi_1(\PP_0)$ has already been thoroughly discussed in~\cite{LevyFmteJ2021}, and the fact that $\Psi_3(\PP_0)$ is a ratio of the other two parameters, we focus on $\Psi_2(\PP_0)$ in this article and propose a new TMLE for it. However, for completeness, $\Psi_1(\PP_0)$ and $\Psi_3(\PP_0)$ will be discussed and serve as companion target parameters whose estimators will also be evaluated via simulations.

\subsubsection{Variance of Treatment Effect}
\noindent To characterize the distribution of the treatment effect, a natural approach is to estimate the mean and the variance of $Y^1 - Y^0$. However, unlike average treatment effect (or ATE) defined as $E(Y^1-Y^0)$, identification of $\mbox{var}(Y^1 - Y^0)$ raises additional challenges and requires additional untestable assumptions~\citep{CoxPE1958, HeckmanMMOP1997, DingDTEVJ2019}. This motivates focus on an alternative parameter $\Psi^F_1(\PP_{O,U})=\mbox{var}(\tau(W))$, which requires none of the additional assumptions. Under the standard identification assumptions listed in previous section we have $\Psi^F_1(\PP_{O,U}) = \Psi_1(\PP_0)$. Then we can re-parametrize it as:
\begin{align*}
	\Psi_1(\PP_0) &\equiv \mbox{var}(\tau_0(W))\\
	&= E(\tau_0^2(W)) - E^2(\tau_0(W))\\
	&= E((\tau_0(W) - ATE)^2)
\end{align*}
Notice that $\Psi_1(\PP_0)$ depends only on $\PP_0(Y|A,W)$ and $\PP_0(W)$. Let $Q_0 \equiv (\bar{Q}_0, Q_{W,0})$, where $\bar{Q}_0, Q_{W,0}$ denote the conditional mean of the outcome and the distribution of covariates respectively. Then we can re-express $\Psi_1(\PP_0)$ as $\Psi_1(Q_0)$. This implies that a substitution estimator can be constructed by only estimating the relevant components $Q_0$ of $\PP_0$ instead of the whole density.\\

\noindent There is an extensive discussion on this VTE parameter in~\cite{LevyFmteJ2021}, where the author proves that it is a pathwise differentiable parameter and derived the efficient influence curve (EIC) of $\Psi_1(Q_0)$ under the nonparametric model:
\begin{align}
    D_{\Psi_1, \PP_0}^* \equiv 2(\tau_0(W) - E(\tau_0(W)))\frac{2A-1}{g_0(A|W)}(Y - \bar{Q}_0(A, W)) + (\tau_0(W) - E(\tau_0(W)))^2 - \Psi_1(Q_0)
\end{align}
where $g_0(A|W)$ = $\PP_0(A|W)$ represent conditional distribution of the treatment, following the notation convention in the targeted learning literature. Influence curves (IC) are model-free, mean-zero functionals that characterise the sensitivity of an estimand to small changes in the data generating law, which can be used for making inference with an asymptotic linear estimator in practice~\citep{HinesVIMHA2022}. EIC is the unique IC living in the tangent space of the statistical model. Notice that, with a little abuse of the notation, the superscript asterisk here is to distinguish the EIC from other ICs instead of marking TMLE-updated terms. Let $R_{\Psi}(\PP_n, \PP_0) \equiv \Psi(Q_n) - \Psi(Q_0) + P_0D^*_{\Psi, \PP_n}$ represent the remainder term, the author of~\cite{LevyFmteJ2021} shows that the remainder term of VTE is 
\begin{align}
 \notag R_{\Psi_1}(\PP_n, \PP_0) &= \Big[E(\tau_0(W))-P_n\tau_n(W)\Big]^2 - E(\tau_0(W)-\tau_n(W))^2 + \\\notag
 &~~~~E\Big\{2(\tau_n(W) - P_n\tau_n(W))\big[\frac{g_0(1|W)-g_n(1|W)}{g_n(1|W)}(\bar{Q}_0(1, W) - \bar{Q}_n(1, W)) - \\
 &~~~~\frac{g_0(0|W)-g_n(0|W)}{g_n(0|W)}(\bar{Q}_0(0, W) - \bar{Q}_n(0, W))\big]\Big\}
\end{align}

\noindent For the last term on the right hand side above, by Cauchy-Schwarz inequality, it can be bounded above by a constant $K$ multiplied by $||\bar{Q}_n - \bar{Q}_{0}||_{L^2(\PP_0)}$ $||g_n - g_0||_{L^2(\PP_0)}$~\citep{LevyFmteJ2021}. So, if $||\bar{Q}_n - \bar{Q}_{0}||_{L^2(\PP_0)}$ $||g_n - g_0||_{L^2(\PP_0)} = o_p(n^{-\frac{1}{2}})$, then the last term will be controlled at rate $o_p(n^{-\frac{1}{2}})$. For the first term, it suffices that that ATE is estimated at rate $o_p(n^{-\frac{1}{2}})$, which holds under the same product rate condition. The second term $E(\tau_0(W)-\tau_n(W))^2$ is more challenging to control as it requires $||\bar{Q}_n - \bar{Q}_{0}||_{L^2(\PP_0)}$ converges at $o_p(n^{-\frac{1}{4}})$. To that end, one can use algorithms such as Highly Adaptive Lasso which achieves fast convergence rate for complex functional forms~\citep{BenkeserHALE2016, vanderLaanADMLJ2023}. More analysis on this remainder term can be found in ~\cite{LevyFmteJ2021}.\\

\subsubsection{Treatment Effect Variable Importance Measure}
 $\Psi_2(\PP_0) \equiv \mbox{var}(\tau_0(W)) - \mbox{var}(\tau_{s,0}(W))$ is proposed in~\cite{HinesVIMHA2022} as a treatment effect VIM. This parameter can be interpreted as a difference in VTEs, quantifying the amount by which the VTE changes when variables in the set $s$ are excluded from the model. More formally, it expresses the additional treatment effect heterogeneity explained by $W_s$, over and above that already explained by $W_{-s}$, where for a vector $u$, we let $u_s$ denote the vector of all components of $u$ with index in $s$~\citep{HinesVIMHA2022}. To distinguish it from another VIM in next section, we call it VIMa. Observe that 
\begin{align}
    \Psi_2(\PP_0) &\equiv \mbox{var}(\tau_0(W)) - \mbox{var}(\tau_{s,0}(W))\\
     &= E(\mbox{var}(\tau_0(W) | W_{-s}))\\
     &= \underbrace{E(
     \underbrace{E(
     \underbrace{\tau_0^2(W)}_{\PP_0(Y|A,W)} |W_{-s})}_{\PP_0(W_{s}|W_{-s})} - \underbrace{E^2(\underbrace{\tau_0(W)}_{\PP_0(Y|A,W)}|W_{-s})}_{\PP_0(W_{s}|W_{-s})})}_{\PP_0(W_{-s})} \\
     &= E(\gamma_{s,0}(W) - \tau_{s,0}^2(W))
\end{align}
where $\gamma_{s,0}(W) \equiv E(\tau_0^2(W)|W_{-s})$. For $O=(W,A,Y)$, the true data generating distribution (DGD) $\PP_0$ can be decomposed as $\PP_0(Y, A, W=(W_s, W_{-s})) = \PP_0(Y|A,W)\PP_0(A|W)\PP_0(W_{s}|W_{-s})\PP_0(W_{-s})$. As equation (5) shows, $\Psi_2(\PP_0)$ depends only on $\PP_0(Y|A,W), \PP_0(W_{s}|W_{-s})$ and $\PP_0(W_{-s})$. Let $\tilde{Q}_0 \equiv (Q_{1,0}, Q_{21,0}, Q_{22,0}, Q_{3,0})$, where $Q_{1,0}$ represent the CATE function $\tau_0$, $Q_{21,0} \mbox{~and~} Q_{22,0}$ represent $\tau_{s,0}$ and $\gamma_{s,0}$ functions respectively, and $Q_{3,0}$ represent the density of $W_{-s}$. These nuisance functions implicitly correspond with $\PP_0(Y|A,W), \PP_0(W_{s}|W_{-s})$ and $\PP_0(W_{-s})$. Then we can re-express the target parameter $\Psi_2(\PP_0)$ as $\Psi_2(\tilde{Q}_0)$. This implies that a substitution estimator can be constructed by sequentially estimating the nuisance functions: $\tau_0$, then conditional mean of $\tau_0$ and $\tau_0^2$ given $W_{-s}$ and finally the mean over $W_{-s}$. In other words, instead of estimating the whole density, we just pursue the conditional means implied by those conditional densities. \\

\noindent Next, we present a decomposition of the EIC which will be useful when the TMLE update step is discussed in the following section. It can be shown that $\Psi_2(\tilde{Q}_0)$ is pathwise differentiable with the EIC $D^*_{\Psi_2, \PP_0} \equiv (\varphi_0(O) - \tau_{s,0}(W))^2 - (\varphi_0(O) - \tau_0(W))^2 - \Psi_2(\tilde{Q}_0)$~\citep{HinesVIMHA2022}, where $\varphi_0(O) = \frac{2A-1}{g_0(A|W)}(Y - \bar{Q}_0(A, W)) + \bar{Q}_0(1, W) - \bar{Q}_0(0, W)$. With some algebra, it can be rewritten as $D^*_{\Psi_2, \PP_0} = 2(\tau_0(W) - \tau_{s,0}(W))\frac{2A-1}{g_0(A|W)}(Y - \bar{Q}_0(A, W)) + (\tau_0(W) - \tau_{s,0}(W))^2 - \Psi_2(\tilde{Q}_0)$. By Lemma 1.6 in section 1.4.5 in~\cite{VanDerLaanUMCL2003}, let $T_{\PP_0}$ denote the tangent space at the distribution $\PP_0$, then we have that $T_{\PP_0} = T_{Y|A,W} \oplus T_{A|W} \oplus T_{W_{s}|W_{-s}} \oplus T_{W_{-s}}$, where $T_{Y|A,W}$, $T_{A|W}$, $T_{W_{s}|W_{-s}}$ and $T_{W_{-s}}$ represent tangent spaces of nuisance parameters. Moreover, let $D^*_{\Psi_2, \PP_0, Y|A,W}, D^*_{\Psi_2, \PP_0, A|W}, D^*_{\Psi_2, \PP_0, W_{s}|W_{-s}}$ and $D^*_{\Psi_2, \PP_0, W_{-s}}$ denote the projections in $L^2(\PP_0)$: $\sqcap(D^*_{\Psi_2, \PP_0}  | T_{Y|A,W})$, $\sqcap(D^*_{\Psi_2, \PP_0} | T_{A|W})$, $  \sqcap(D^*_{\Psi_2, \PP_0} | T_{W_s|W_{-s}})$ and $\sqcap(D^*_{\Psi_2, \PP_0} | T_{W_{-s}})$ respectively. Then, the EIC can be orthogonally decomposed as $D^*_{\Psi_2, \PP_0} = D^*_{\Psi_2, \PP_0, Y|A,W} + D^*_{\Psi_2, \PP_0, A|W} + D^*_{\Psi_2, \PP_0, W_{s}|W_{-s}} + D^*_{\Psi_2, \PP_0, W_{-s}}$. And it can be shown that $D^*_{\Psi_2, \PP_0, Y|A,W} = 2(\tau_0(W) - \tau_{s,0}(W))\frac{2A-1}{g_0(A|W)}(Y - \bar{Q}_0(A, W))$, $D^*_{\Psi_2, \PP_0, A|W}=0$, $D^*_{\Psi_2, \PP_0, W_{s}|W_{-s}} = \tau_0^2(W) - E(\tau_0^2(W)|W_{-s}) - 2\tau_{s,0}(W)(\tau_0(W) - \tau_{s,0}(W))$ and $D^*_{\Psi_2, \PP_0, W_{-s}}= E(\tau_0^2(W)|W_{-s}) - \tau_{s,0}^2(W) - \Psi_2(\tilde{Q}_0)$. The details of the decomposition can be found in Appendix. This decomposition result of EIC will be useful when we discuss TMLE update step in following section.\\

\noindent Similar to $\Psi_1(Q_0)$, the remainder term of $\Psi_2(\tilde{Q}_0)$ can be represented by:
\begin{align}
    \notag R_{\Psi_2}(\PP_n, \PP_0) 
    ={}&E\Big[(\tau_{s,0}(W) - \tau_{s,n}(W))^2- (\tau_0(W) - \tau_n(W))^2\Big] + \\\notag
    &{} E\Big\{2(\tau_n(W) - \tau_{s,n}(W))[\frac{g_0(1|W)-g_n(1|W)}{g_n(1|W)}(\bar{Q}_0(1, W) - \bar{Q}_n(1, W)) - \\
    &{}\frac{g_0(0|W)-g_n(0|W)}{g_n(0|W)}(\bar{Q}_0(0, W) - \bar{Q}_n(0, W))]\Big\}
\end{align}

\noindent Observe that $R_{\Psi_2}(\PP_n, \PP_0)$ can be decomposed into the two terms. The second term is a standard ``double robust'' term which can be bounded above by a constant multiplied by $||\bar{Q}_n - \bar{Q}_{0}||_{L^2(\PP_0)}||g_n - g_0||_{L^2(\PP_0)}$. This is similar to the last term of $R_{\Psi_1}(\PP_n, \PP_0)$. Then, the challenging part is to control the first term $E((\tau_{s,0}(W) - \tau_{s,n}(W))^2- (\tau_0(W) - \tau_n(W))^2)$ since the integrand in the expectation is a difference of squares. To make sure this term is $o_p(n^{-\frac{1}{2}})$, we need that $\tau_0 - \tau_n$ and $\tau_{s,0} - \tau_{s,n}$ are both $o_p(n^{-\frac{1}{4}})$ in $L^2(\PP_0)$ norm. This means, even in randomized controlled trial where the ``double robust'' remainder term can be eliminated, we still need good CATE estimates to control the first remainder term, thereby guarantee the performance of the estimator. The details of the derivation of the remainder term can be found in Appendix. 

\subsubsection{Scaled Treatment Effect Variable Importance Measure}
\noindent Finally, we consider a scaled VIM $\Psi_3(\PP_0) \equiv \Psi_2(\tilde{Q}_0) / \Psi_1(Q_0) = 1 - \frac{\mbox{var}(\tau_{s,0}(W))}{\mbox{var}(\tau_0(W))}$. Refer it as VIMb for convinience. This parameter is essentially VIMa scaled by VTE. It has the same interpretation as VIMa, with additional convenience when researchers want to compare the relative importance of covariate sets $s$ and $s'$ by comparing the magnitudes their VIMb estimates~\citep{HinesVIMHA2022}. Let $\breve{Q}_0 = (Q_0, \tilde{Q}_0)$, where $\breve{Q}_0$ contains all the relevant components. Specifically, it includes $\bar{Q}_0$ (the conditional mean of the outcome), $Q_{W,0}$ (the distribution of the covariates), $Q_{1,0}$ (CATE), $Q_{21,0}$ (conditional mean of CATE given $W_{-s}$), and $Q_{22,0}$ (conditional mean of squared CATE given $W_{-s}$). Now, we can re-express $\Psi_3(\PP_0)$ with $\Psi_3(\breve{Q}_0)$. Applying the delta method, one can show the EIC of this parameter is $D_{\Psi_3, \PP_0}^* = (D_{\Psi_2, \PP_0}^* - \Psi_3(\breve{Q}_0)D_{\Psi_1, \PP_0}^*)/\Psi_1(Q_0)$~\citep{HinesVIMHA2022}. Without deriving the explicit form of the remainder term, we know that it is controlled by the remainders of $\Psi_1(Q_0)$ and $\Psi_2(\tilde{Q}_0)$.  We omit the analysis of the remainder term here and leave it for future investigation.

\section{TMLE for Treatment Effect VIM}
\subsection{Overview of TMLE}
A plug-in estimator (also called substitution estimator) applies the target parameter mapping directly to an estimate of relevant components. For example, a simple substitution estimator of ATE is $\Psi(Q_n^0) = \frac{1}{n} \sum_{i=1}^{n} (\bar{Q}_n^{0}(1,w_i) - \bar{Q}_n^{0}(0,w_i))$, where the relevant components are the conditional mean of the outcome and the marginal distribution of the covariates (same as VTE).  The advantage of a plug-in estimator is that it always respects the global constraints on the data distribution and that the estimand is a particular function of the distribution. A targeted maximum likelihood
estimator is a plug-in estimator that generally consists of two steps: 1) obtain an initial estimate $\PP_n^{0}$ of the data-generating distribution $\PP_0$ (or the relevant portion $Q_0$ of $\PP_0$), and 2) update this initial fit. It uses a path through $\PP_n^{0}$ the canonical gradient and computes the TMLE update along that path through the maximum likelihood estimator of the fluctuation parameter~\citep{LaanTLCI2011}. The second step in practice often contains iterative update steps and the stopping criterion is that $P_nD_{\Psi_2, \PP_n^*}^* = 0$ is solved up till desired approximation ($o_p(n^{-\frac{1}{2}})$), where $D_{\Psi_2, \PP_n^*}^*$ represents the EIC at the TMLE-updated fit $\PP_n^*$.

\subsection{Initial Estimation of $\Psi_2(\tilde{Q}_0)$}
In this section, we introduce two algorithms for the general initial estimation of $\Psi_2(\tilde{Q}_0)$, and two metalearner methods for CATE estimation. We also give an example on how to use these methods to obtain initial estimates of the nuisance functions and construct a simple substitution estimate for $\Psi_2(\tilde{Q}_0)$.

\subsubsection{Machine Learning Algorithms}
For the initial estimation, we consider Super Learner (SL) and/or Highly Adaptive Lasso (HAL), which often serve as “off the shelf” algorithms in targeted learning.
SL~\citep{LaanSLS2007} is a loss-based ensemble learning method that selects from a group of candidate algorithms by minimizing the risk via K-folds cross-validation and assigning weights to individual algorithms based on their performance. The discrete SL can be used to select the best prediction algorithm from among a supplied library of machine learning algorithms. The ensemble SL can be used to assign weights to the library of candidate learners so as to create a combination of these learners that minimizes the cross-validated risk with respect to an appropriate loss function. This notion of weighted combinations has also been referred to as stacked regression~\citep{BreimanSRJ1996}. SL provides an important step in creating a robust estimator.  
The other algorithm we consider is HAL. It is a nonparametric loss based estimator, which is capable of estimating complex functional parameters with mild assumptions that the true functional parameter is right-hand continuous with left-hand limits and has variation norm smaller than a constant, but neither relies upon local smoothness assumptions nor is constructed using local smoothing techniques~\citep{BenkeserHALE2016}. 
The HAL estimator can be represented in the following form~\citep{BenkeserHALE2016, vanderLaanEEPDJ2021}:
\begin{align}
    \hspace{1.5cm} \psi_{\beta} = \beta_0 + \sum_{s\subset \{1,2,...,p\}}\sum_{i=1}^{n} \beta_{s,i} \phi_{s,i} \hspace{0.5cm}\mbox{with}\hspace{0.5cm} \beta_0 + \sum_{s\subset \{1,2,...,p\}}\sum_{i=1}^{n} \big | \beta_{s,i} \big | < C
\end{align}
where $s$ denotes any subset of $\{1,2,...,p\}$, and $\phi_{s,i}:  x_s \mapsto I(\tilde{x}_{s,i} \leq x_s)$ are the indicator basis functions defined by the support points $\tilde{x}_{s,i}$ from the observations. In other words, the HAL estimator constructs a linear combination of indicator basis functions to minimize the loss-specific empirical risk under the constraint that the $L_1$ norm of the vector of coefficients is bounded by a finite constant matching the sectional variation norm of the target functional~\citep{ErtefaieNIPWJ2021}. HAL allows for arbitrarily complex functional forms to be estimated at convergence rates $o_p(n^{-\frac{1}{3}})$ up till $\log{n}$ factor and has been shown to have competitive finite-sample performance relative to many other popular machine learning algorithms~\citep{BenkeserHALE2016, BibautFRERA2019, vanderLaanEEPDJ2021}.

\subsubsection{CATE Metalearners}

CATE is an important intermediate term for estimating VTE and VIM parameters. In this article, we consider two CATE metalearners. By the definition of the CATE, the direct method is to estimate the conditional expectation of the outcome and then take the difference between $\bar{Q}_n(1,W)$ and $\bar{Q}_n(0,W)$. This yields the first CATE estimator and we call it the S-learner~\citep{KunzelMEHTM2019}. Alternatively, recall that we defined the pseudo outcome as $\varphi(O) = \frac{2A-1}{g(A|W)}(Y - \bar{Q}(A, W)) + \bar{Q}(1, W) - \bar{Q}(0, W)$. It can be shown that the pseudo outcome $\varphi(O)$ is equivalent to CATE in expectation (i.e. conditional mean of $\varphi(O)$ given $W$ equals $\tau(W)$). Thus, an alternative CATE estimate can be obtained by regressing the pseudo outcome estimate $\varphi_n(O)$ on the covariates $W$~\cite{vanderLaanTLODO2013, LuedtkeSODTM2016a, KennedyODREA2020}. We call this estimator the DR-learner. 

\subsubsection{Estimation of the Nuisance Functions}

As it was mentioned earlier, the nuisance functions associated with $\Psi_2(\tilde{Q}_0)$ include $\tau_0$, $\tau_{s,0}$, $\gamma_{s,0}$, and finally the mean over $W_{-s}$. Notice the last one will be handled by the taking the empirical mean. So we do not need any estimation on that one. Then, once we decide which algorithms and CATE metalearners to use, we can then estimate the nuisance functions and then construct an initial estimate of $\Psi_2(\tilde{Q}_0)$.  For example, if one chooses HAL and S-learner, then one should: 1) fit a HAL model for $\bar{Q}_0$; 2) take the difference between $\bar{Q}_n^{(0)}(1,W)$ and $\bar{Q}_n^{(0)}(0,W)$ to get $\tau_n^{(0)}(W)$, an initial S-learner estimate of CATE; 3) fit another two HAL models, with $\tau_n^{(0)}(W)$ and $(\tau_n^{(0)}(W))^2$ as outcomes respectively and $W_{-s}$ as covariates. Then we get $\tau_{s,n}^{(0)}(W)$ and $\gamma_{s,n}^{(0)}(W)$.
Alternatively, if one chooses DR-learner, then he needs additionally fit a HAL model for $g_0$ at step 1 above (if $g_0$ is unknown). Then at step 2, one needs compute $\varphi_n^{(0)}(O)$ with previous estimates of $\bar{Q}_0$ and $g_0$. Then using $\varphi_n^{(0)}(O)$ as the outcome and $W$ as covariates to fit a HAL model and then obtain a initial DR-learner estimate of CATE. The estimation method discussed uses full data for algorithm training. In practice, If overfitting is a concern, for example because of using aggressive nonparametric machine learning (ML) algorithms, the sample splitting scheme proposed in \cite{HinesVIMHA2022} can be adopted.\\

\noindent With the estimates of these nuisance functions, we can now construct a simple substitution estimate for $\Psi_2(\tilde{Q}_0)$. As the definition in equation (6) shows, we know that a SS estimator can be defined as:
\begin{align}
     &\Psi_2(\tilde{Q}_n^0) \equiv \frac{1}{n}\sum_{i=1}^n [\gamma_{s,n}^{(0)}(w_i) - (\tau_{s,n}^{(0)}(w_i))^2]
\end{align}

\noindent So we can simply plug into formula (10) the estimates from step 3 above. Now we have a SS estimate for $\Psi_2(\tilde{Q}_0)$. To obtain a TMLE, one needs to further update the initial estimates of the nuisance functions above in an iterative manner until $P_nD_{\Psi_2, \PP_n^*}^* \approx 0$. 

\subsection{Targeted Maximum Likelihood Estimation of $\Psi_2(\tilde{Q}_0)$}

\noindent Now, for $\Psi_2(\tilde{Q}_0)$, we propose a new plug-in TMLE defined as follows. 
\begin{align}
     &\Psi_2(\tilde{Q}_n^*) \equiv \frac{1}{n}\sum_{i=1}^n [\gamma_{s,n}^*(w_i) -  (\tau_{s,n}^*(w_i))^2]
\end{align}
where $\gamma_{s,n}^*$ and $\tau_{s,n}^*$ are the TMLE-updated terms. For this TMLE, we have the following extended theorem from targeted learning literature~\citep{LaanTLCI2011, VanDerLaanTLDS2018}:

\begin{theorem}
Consider $O \sim{\PP_0} \in \mathcal{M}$. Let $\Psi_2(\tilde{Q}_n^*): \mathcal{M} \rightarrow \RR$ be defined by $\Psi_2(\tilde{Q}_n^*) \equiv \frac{1}{n}\sum_{i=1}^n [\gamma_{s,n}^*(w_i) -  (\tau_{s,n}^*(w_i))^2]$. Let $D^*_{\Psi_2, \PP_0}$ and $D^*_{\Psi_2, \PP_n^*}$ be the efficient influence curves of parameter mapping $\Psi_2$ at $\PP_0$ and $\PP_n^*$ respectively. Let $R_{\Psi_2}(\PP_n^*, \PP_0)$ be defined as $R_{\Psi_2}(\PP_n^*, \PP_0) \equiv \Psi_2(\tilde{Q}_n^*) - \Psi_2(\tilde{Q}_0) + P_0 D_{\Psi_2, \PP_n^*}^*$. Assume the following TMLE conditions:\\
\textit{A1.} $L^2$ consistency: $||D_{\Psi_2, \PP_n^*}^*-D_{\Psi_2, \PP_0}^*||_{L^2(\PP_0)} \overset{P}{\rightarrow} 0$.\\
\textit{A2.} Donsker condition: $D_{\Psi_2, \PP_n^*}^*$ falls into a Donsker class with probability tending to 1. \\
\textit{A3.} $R_{\Psi_2}(\PP_n^*, \PP_0)$ is $o_p(n^{-\frac{1}{2}})$.\\
\textit{A4.} $P_nD_{\Psi_2, \PP_n^*}^*$ is $o_p(n^{-\frac{1}{2}})$.\\

\noindent Then $\Psi_2(\tilde{Q}_n^*)$ is a an asymptotically efficient estimator of $\Psi_2(\tilde{Q}_0)$:

$$\Psi_2(\tilde{Q}_n^*) - \Psi_2(\tilde{Q}_0) = P_n D_{\Psi_2, \PP_0}^* + o_p(n^{-\frac{1}{2}}).$$
\end{theorem}

\noindent More discussion on Theorem 1 can be found in Appendix. From the literature we also have that $\Psi_2(\tilde{Q}_n^*) - \Psi_2(\tilde{Q}_0) = P_n D^*_{\Psi_2, \PP_0} + (P_n - P_0)(D_{\Psi_2, \PP_n^*}^*-D^*_{\Psi_2, \PP_0}) - P_n D_{\Psi_2, \PP_n^*}^* + R_{\Psi_2}(\PP_n^*, \PP_0)$~\citep{LaanTLCI2011, VanDerLaanTLDS2018}. To prove the theorem, we need to show that $(P_n - P_0)(D_{\Psi_2, \PP_n^*}^*-D^*_{\Psi_2, \PP_0}) - P_n D_{\Psi_2, \PP_n^*}^* + R_{\Psi_2}(\PP_n^*, \PP_0)$ is $o_p(n^{-\frac{1}{2}})$. One sufficient condition is that each term is $o_p(n^{-\frac{1}{2}})$. It turns out that: 1) A1 and A2 guarantees that $(P_n - P_0)(D_{\Psi_2, \PP_n^*}^*-D^*_{\Psi_2, \PP_0})$ is $o_p(n^{-\frac{1}{2}})$, 2) by using HAL or SL with HAL as candidate, we can make A3 hold~\citep{LaanTLCI2011, VanDerLaanTLDS2018}, and 3) TMLE arranges A4~\citep{LaanTLCI2011, VanDerLaanTLDS2018}. When the statement $\Psi_2(\tilde{Q}_n^*) - \Psi_2(\tilde{Q}_0) = P_n D^*_{\Psi_2, \PP_0} + o_p(n^{-\frac{1}{2}})$ holds, $\Psi_2(\tilde{Q}_n^*) $ converges to $\Psi_2(\tilde{Q}_0)$ in probability, and $\sqrt{n}(\Psi_2(\tilde{Q}_n^*) - \Psi_2(\tilde{Q}_0)) \overset{D}{\Rightarrow} \mathcal{N}(0, \mbox{var}(D^*_{\Psi_2, \PP_0}))$. \\

\noindent As it was discussed previously, the EIC of $\Psi_2(\tilde{Q}_0)$ can be orthogonally decomposed as $D^*_{\Psi_2, \PP_0} = D^*_{\Psi_2, \PP_0, Y|A,W} + D^*_{\Psi_2, \PP_0, W_{s}|W_{-s}} + D^*_{\Psi_2, \PP_0, W_{-s}}$. For each initial estimator of a nuisance parameter, the TMLE uses a path through this estimator whose score (w.r.t. appropriate loss) equals the corresponding EIC component. In that manner, the TMLE update of the nuisance parameters will solve the corresponding score equations and thereby solve also the EIC score equation. In other words, the TMLE update is constructed to solve each of $P_nD^*_{\Psi_2, \PP_n^*, Y|A,W} = 0, P_nD^*_{\Psi_2, \PP_n^*, W_{s}|W_{-s}}= 0~\mbox{and}~P_nD^*_{\Psi_2, \PP_n^*, W_{-s}}= 0$) at $\PP_n^*$. Thereby the whole EIC score equation ($P_nD_{\Psi_2, \PP_n^*}^*=0$) is solved. In practice, each component score equation is solved by carefully choosing a parametric fluctuation model that approximates a universal least favourable model~\citep{LaanTLCI2011, VanDerLaanTLDS2018}. There are often more than one way of constructing such fluctuation model in reality. In addition, one notices that the $D^*_{\Psi_2, \PP_n, W_{-s}}$ term will be automatically solved by taking the empirical mean so we do not need to do TMLE update for it. The implementation of this TMLE can be found below (also see pseudo code in Algorithm 1).\\

\noindent \textbf{TMLE alogrithm for $\Psi_2(\tilde{Q}_0)$ in words}
\begin{enumerate}
    \item First use HAL or SL to obtain initial estimates: $\bar{Q}_n^{_{(0)}}(A,W), g_n(A|W)$, $\tau_n^{_{(0)}}(W), \tau_{s,n}^{_{(0)}}(W)$ and $\gamma_{s,n}^{_{(0)}}(W)$. Then, just as we sequentially get the initial estimates, we will also target them in a specific order.
    \item Recall that the goal is to solve $P_n D^*_{\Psi_2, \PP_n^*} \approx 0$. And we know that $P_n D^*_{\Psi_2, \PP_n^*}= P_n( D^*_{\Psi_2, \PP_n^*, Y|A,W} + D^*_{\Psi_2, \PP_n^*, W_{s}|W_{-s}} + P_nD^*_{\Psi_2, \PP_n^*, W_{-s}})$. So we will tackle this problem in a divide-and-conquer manner. In practice, we split $D^*_{\Psi_2, \PP_n^*, W_{s}|W_{-s}}$ into $2\tau_{s,n}(W)(\tau_n(W) - \tau_{s,n}(W))$ and $\tau_n^2(W) - E(\tau_n^2(W)|W_{-s})$. To simplify the notation, we denote them with $D_{1,W_{s}|W_{s}, n}$ and $D_{2,W_{s}|W_{s}, n}$. Also, we use $D_{Y|A,W,n}$ as an alternative notation for $D^*_{\Psi_2, \PP_n^*, Y|A,W}$. In addition, we compute ``clever covariates'' $H_{1,n} = 2(\tau_n(W) - \tau_{s,n}(W))\frac{2A-1}{g_n(A|W)}$, $H_{1,q,n} = 2(\tau_n(W) - \tau_{s,n}(W))(2A-1)$ and $H_{2,n} = 2\tau_{s,n}(W)$. These will be used to construct the fluctuation models such that the scores equal the EIC components.
    Now, we update the initial estimates in the order of i) $\bar{Q}_n$ and $\tau_n$, ii) $\tau_{s,n}$ and iii) $\gamma_{s,n}$ through three linear fluctuation models: a) $\bar{Q}_n^{_{(i)}}(A,W) = \bar{Q}_n^{_{(i-1)}}(A,W) + \epsilon_1 H_{1,q,n}^{_{(i-1)}} P_n(D_{Y|A,W,n}^{_{(i-1)}})$; b) $ \tau_{s,n}^{_{(i)}}(W) = \tau_{s,n}^{_{(i-1)}}(W) + \epsilon_2 H_{2,n}^{_{(i-1)}}P_n(D_{1,W_{s}|W_{s}, n}^{_{(i-1)}})$; c) $ (\tau^*_n(W))^2 \sim \mbox{offset}(\gamma_{s,n}^{_{(0)}}(W)) + \epsilon_3$.
    \item For nuisance function i) and ii), the targeting step is an iterative procedure since the clever covariates contain the updated terms. We calculate the clever covariate $H_n$ and the estimated EIC component (e.g. $D_{Y|A,W}^*$). We will update the initial estimates toward the direction that maximally decreases the loss (i.e. the product of the clever covariate and the empirical mean of the estimated EIC)~\citep{LaanTLCI2011, VanDerLaanTLDS2018}. We first update $\bar{Q}_n$ with small step size $\epsilon_1$, and get updated $\tau_n$ by taking the difference between $\bar{Q}_n(1,W)$ and $\bar{Q}_n(0,W)$. Then we update $\tau_{s,n}$ with small step size $\epsilon_2$. After one round of update, we recompute the clever covariates and the estimated EICs. Then we check whether the empirical mean of the estimated EIC is solved at a proper level. If not, we repeat this step until 1) and 2) are both solved. Finally we get $\bar{Q}_n^*$, $\tau_n^*$ and $\tau_{s,n}^*$.
    \item For nuisance function iii), the clever covariate is just a constant so we can simply fit a parametric linear regression and solve 3) at a single step. Then we extract the $\epsilon_3$ from the model fit and use it to get the TMLE-updated $\gamma_{s,n}^*$.
    \item Now we have solved the whole score equation $P_n D^*_{\Psi_2, \PP_n^*} \approx 0$ by solving each component. Plug $\tau_{s,n}^{_{*}}(W)$ and $\gamma_{s,n}^{_{*}}(W)$ into the formula (11) and obtain the TMLE.
\end{enumerate}

\begin{algorithm}[!ht]
\SetAlgoLined
 Fit $\bar{Q}_n^{_{(0)}}(A,W), g_n(A|W), \tau_n^{_{(0)}}(W), \tau_{s,n}^{_{(0)}}(W)$ and $\gamma_{s,n}^{_{(0)}}(W)$.\\[2pt]
 $H_{1,n}^{_{(0)}} = 2(\tau_n^{_{(0)}}(W) - \tau_{s,n}^{_{(0)}}(W))\frac{2A-1}{g_n(A|W)}$; 
 $H_{1,q,n}^{_{(0)}} = 2(\tau_n^{_{(0)}}(W) - \tau_{s,n}^{_{(0)}}(W))(2A-1)$\\[2pt]
 $D_{Y|A,W,n}^{_{(0)}} = H_{1,n}^{_{(0)}}(Y - \bar{Q}_n^{_{(0)}}(A, W))$.\\[2pt]
 $\hat\sigma_{1, n} = \mbox{sd}(D_{Y|A,W,n}^{_{(0)}})$\\[2pt]
 $H_{2,n}^{_{(0)}} = 2\tau_{s,n}^{_{(0)}}(W)$.\\[2pt]
 $D_{1,W_{s}|W_{s}, n}^{_{(0)}} = H_{2,n}^{_{(0)}}(\tau_n^{_{(0)}}(W) - \tau_{s,n}^{_{(0)}}(W))$.\\[2pt]
 $\hat\sigma_{2, n} = \mbox{sd}(D_{1,W_{s}|W_{s}, n}^{_{(0)}})$\\[2pt]
 $\epsilon_1 = \epsilon_2 = 0.0001$ \\[2pt]
 $i = 1$\\[2pt]
 \While{True}{
 \vspace{0.1cm}
 Step 1. Update $\bar{Q}_n$ and $\tau_n$.
 \begin{align*}
     \bar{Q}_n^{_{(i)}}(A,W) &= \bar{Q}_n^{_{(i-1)}}(A,W) + \epsilon_1 H_{1,q,n}^{_{(i-1)}} P_n(D_{Y|A,W,n}^{_{(i-1)}}) \\
     \tau_n^{_{(i)}}(W) &= \bar{Q}_n^{_{(i)}}(1,W) - \bar{Q}_n^{_{(i)}}(0,W)
 \end{align*}
 
 Step 2. Update $\tau_{s,n}(W)$.
 $$ \tau_{s,n}^{_{(i)}}(W) = \tau_{s,n}^{_{(i-1)}}(W) + \epsilon_2 H_{2,n}^{_{(i-1)}}P_n(D_{1,W_{s}|W_{s}, n}^{_{(i-1)}})$$
 
 Step 3. Check the criterion.\\[2pt]
 $H_{1,n}^{_{(i)}} = 2(\tau_n^{_{(i)}}(W) - \tau_{s,n}^{_{(i)}}(W))\frac{2A-1}{g_n(A|W)}$.\\[2pt]
 $D_{Y|A,W,n}^{_{(i)}} = H_{1,n}^{_{(i)}}(Y - \bar{Q}_n^{_{(i)}}(A, W))$.\\[2pt]
 $H_{2,n}^{_{(i)}} = 2\tau_{s,n}^{_{(i)}}(W)$.\\[2pt]
 $D_{1,W_{s}|W_{s}, n}^{_{(i)}} = H_{2,n}^{_{(i)}}(\tau_n^{_{(i)}}(W) - \tau_{s,n}^{_{(i)}}(W))$.\\[2pt]
 $c_1 \leftarrow |P_n(D_{Y|A,W,n}^{_{(i)}})| \leq \frac{\hat\sigma_{1, n}}{\sqrt{n}\log{n}}$\\[2pt]
 $c_2 \leftarrow |P_n(D_{1,W_{s}|W_{s}, n}^{_{(i)}})| \leq \frac{\hat\sigma_{2, n}}{\sqrt{n}\log{n}}$\\[2pt]
  \uIf{$c_1$ and $c_2$}{
  \vspace{0.1cm}
  $\bar{Q}^*_n(A,W) = \bar{Q}_n^{_{(i)}}(A,W)$; 
  $\tau^*_n(W) = \tau_n^{_{(i)}}(W)$;
  $\tau_{s,n}^*(W) = \tau_{s,n}^{_{(i)}}(W)$\\
  \textbf{break}
  }
  $i = i + 1$
 }
 Update $\gamma_{s,n}(W)$ by a linear regression:
 $$ (\tau^*_n(W))^2 \sim \mbox{offset}(\gamma_{s,n}^{_{(0)}}(W)) + \epsilon_3$$\\[2pt]
 Obtain $\gamma_{s,n}^*$ from the update above.\\[2pt]
 $\Psi_2(\tilde{Q}_n^*)  = \frac{1}{n}\sum_{i=1}^n \{ \gamma_{s,n}^*(w_{i}) -  (\tau_{s,n}^*(w_{i}))^2\}$\\[2pt]
 
\caption{TMLE for $\Psi_2(\tilde{Q}_0)$}
\end{algorithm}

\noindent There are a few changes made due to practical considerations for this TMLE algorithm, compared with the classical TMLE procedure. For this TMLE, we use linear fluctuation model instead of the logistic model that is used in most previous TMLE implementations. The motivation is that we want to circumvent the tedious variable transformations involved in the logistic TMLE update steps. To use logistic fluctuation here, one must first transform the outcome variable $Y$ into $[0,1]$. Then one also needs to make sure the $\tau_n(W)$ and $\tau_{s,n}(W)$ terms are bounded between $[0,1]$ for the logistic update, which requires additional variable transformations in the \texttt{while} loop. However, one well-known caveat of using a linear TMLE update is that it might extrapolate the variable outside its global bound when the propensity score term inside the ``clever covariates'' (in this case, $H_{1,n}^{_{(0)}}$) is extreme. To handle this, we separate the $1/g_n(A|W)$ term from the clever covariates and use it as the weights in the loss function of the linear model.

\subsection{Other Estimators}
\noindent The SS, EE and TMLE estimators for $\Psi_1(Q_0)$ can be found in \cite{LevyFmteJ2021}. For $\Psi_2(\tilde{Q}_0)$, the EE estimator can be computed with $\frac{1}{n}\sum_{i=1}^{n} \{(\varphi_n(o_i) - \tau_{s,n}(w_i))^2 - (\varphi_n(o_i) - \tau_n(w_i))^2\}$ as proposed in~\cite{HinesVIMHA2022}. For $\Psi_3(\breve{Q}_0)$, we can simply construct the SS, EE and TMLE estimators by plugging into their $\Psi_1(Q_0)$ and $\Psi_2(\tilde{Q}_0)$ counterparts. We need to clarify that the SS estimator here is not a completely plug-in estimator if one estimates $\Psi_1(Q_0)$ and $\Psi_2(\tilde{Q}_0)$ separately, because $Q_n$ and $\tilde{Q}_n$ are not necessarily compatible in this case. Similarly, the TMLE, by directly taking the ratio of $\Psi_2(\tilde{Q}_n^*)$ and $\Psi_1(Q_n^*)$, is not a pure plug-in TMLE. A plug-in TMLE would require solving $P_nD_{\Psi_3, \PP_n^*}^*$ directly by decomposing the EIC and constructing fluctuation models to solve each score components (just as we did for $\Psi_2(\tilde{Q}_0)$). In this article, we still name these two kind of estimators SS and TMLE since 1) they are both valid estimators based on SS and TMLE methods, and 2) asymptotically they should have similar behaviour to the real SS and TMLE. As for inference, the CIs of these estimators can be computed with the empirical variance of the estimated EIC. \\

\noindent Besides the estimating methods mentioned above, we also think it is worth exploring a pure plug-in HAL estimator. Recent papers show that plug-in HAL can be efficient when we undersmooth it, and sometimes achieve super efficiency without undersmoothing~\citep{vanderLaanEEPDJ2021, vanderLaanADMLJ2023}. Moreover, the recent developed Poisson HAL method for intensities can be used to directly estimate the conditional densities such as $\PP_0(W_{s}|W_{-s})$~\citep{RytgaardEtieM2023}. So, we think a plug-in estimator with HAL to estimate all relevant components (even conditional densities) could be good enough, even without a targeting step. Due to time constraint, we leave this for future investigation.

\section{Simulation}
\subsection{Simulation Setting}
We describe our simulation in terms of the sample size $n$, the dimension of covariates $p$, as well as the following data generating distributions (DGD):
\begin{eqnarray*}
W_{1} &\sim& Unif(-1, 1)\\[1pt]
W_{2} &\sim& Unif(-1, 1)\\[1pt]
A &\sim& Bernoulli(p) \mbox{~~where~~} p = \mbox{expit}(0.1*W_1*W_2-0.4*W_1) \\[1pt]
\tau &=& W_1^2*(W_1+7/5) + (5*W_2/3)^2\\[1pt]
\mu_{Y} &=& A*\tau + W_1*W_2 + 2*W_2^2 - W_1 \\[1pt]
Y &\sim& N(\mu_{Y}, 1)
\end{eqnarray*}

\noindent As the DGD above shows, we have two independent continuous covariates and an outcome variable with homoscedastic noise. For each target parameter (VTE, VIMa and VIMb), we evaluate all three types of estimators (SS, EE and TMLE) in the simulation. We use HAL (with default arguments in the \texttt{hal9001}~\citep{coyle2021hal9001-rpkg, HejaziHSHAS2020} package) to estimate the nuisance functions. In addition, we implement both S-learner and DR-learner for CATE estimation. We find the relative performance of estimators with these two methods are similar in this simulation. Consequently, we present the results with S-learner below and attach the results with DR-learner in the Appendix. 
In this simulation, we consider the set of experiment sample sizes $\{2e2, 5e2, 1e3, 3e3, 5e3, 1e4, 2e4\}$. For each sample size value, $500$ iterations are executed to collect the point estimates and CIs for each estimator. We truncate the estimated propensity scores between $[0.025, 0.975]$ to alleviate potential positivity violations. \\


\noindent We implemented our simulations in R~\citep{R_general}, using the packages \texttt{hal9001}~\citep{coyle2021hal9001-rpkg, HejaziHSHAS2020} and \texttt{sl3}~\citep{coyle2021sl3} for estimation. Details of the implementation, including the code, can be found at
\url{github.com/HaodongL/te_vim}.\\

\subsection{Estimators' Performance}
We arrange the simulation results of the three target parameters in the order of VTE, VIMa and VIMb. For each parameter, we present the same set of performance metrics: mean squared error (MSE), absolute value of bias, coverage of 0.95 confidence intervals (CI), and oracle coverage of CIs. The difference between the two coverage metrics is that the former is calculated with estimated variance of EIC, whereas the latter is calculated with the true sampling variance of each estimator across 500 iterations.  \\

\subsubsection{Results on the VTE Parameter} 
\begin{figure}[ht]
    \centering
    \includegraphics[scale = 0.35]{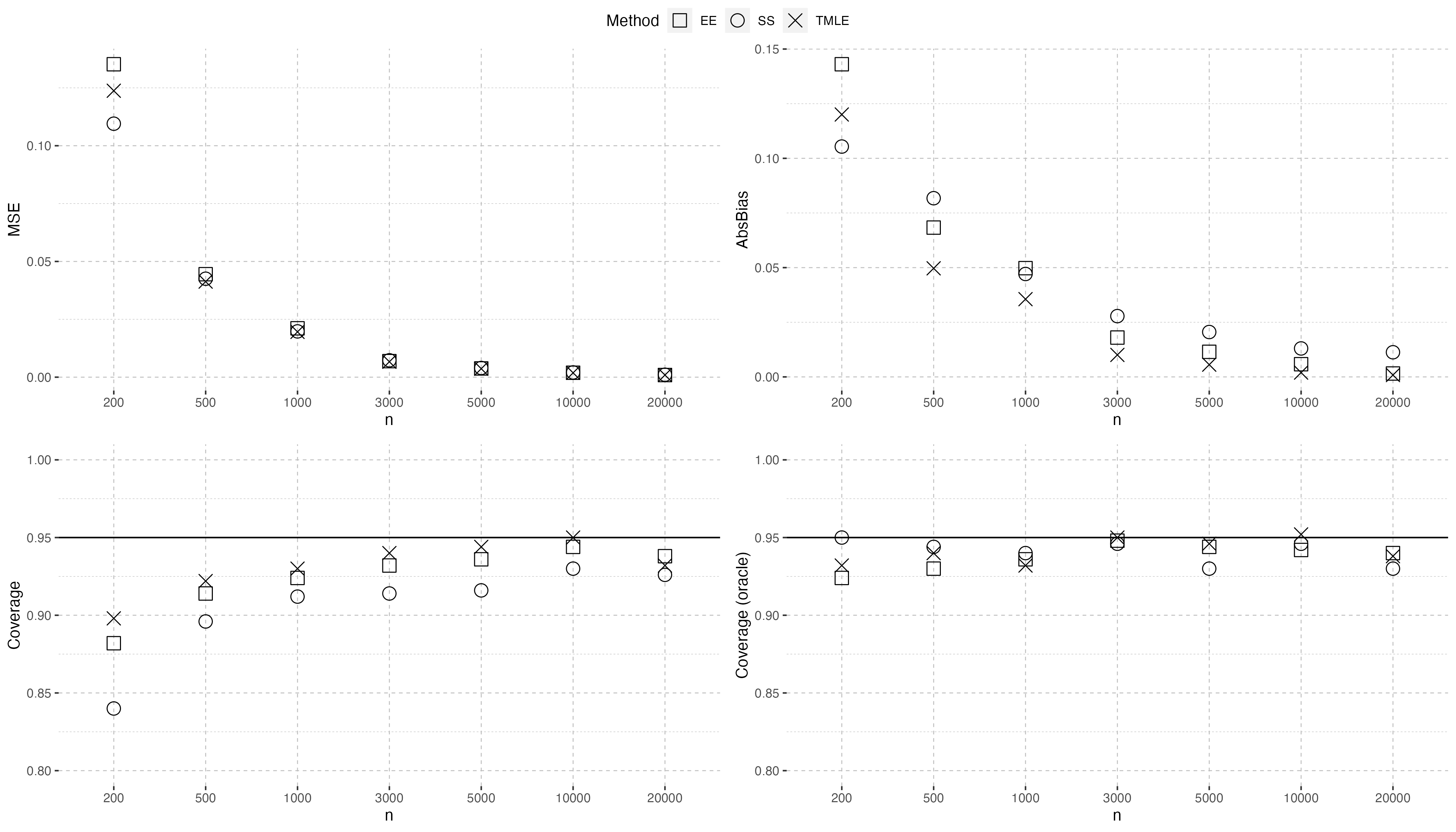}
    \caption{Main performance metrics of the estimators of the VTE parameter (Squares circles and crosses represent EE, SS and TMLE respectively. For the two metrics in the first row, the lower the value, the better. For the two coverage metrics on the second row, the closer to the 0.95 reference line, the better.)}
    \label{p1}
\end{figure}

\noindent For the VTE parameter $\Psi_1(Q_0)$, Figure~\ref{p1} shows that the bias and MSE decrease and the coverage increases as sample size increases for all three estimators. TMLE outperforms EE and SS estimators in terms of lower bias and better coverage across almost all sample sizes (When sample size is $200$, the SS estimator has smaller bias but worse coverage). Three estimators have similar MSE. And their oracle coverages are all close to $0.95$.

\newpage
\subsubsection{Results on the Unscaled VIM Parameter} 

\begin{figure}[ht]
    \centering
    \includegraphics[scale = 0.35]{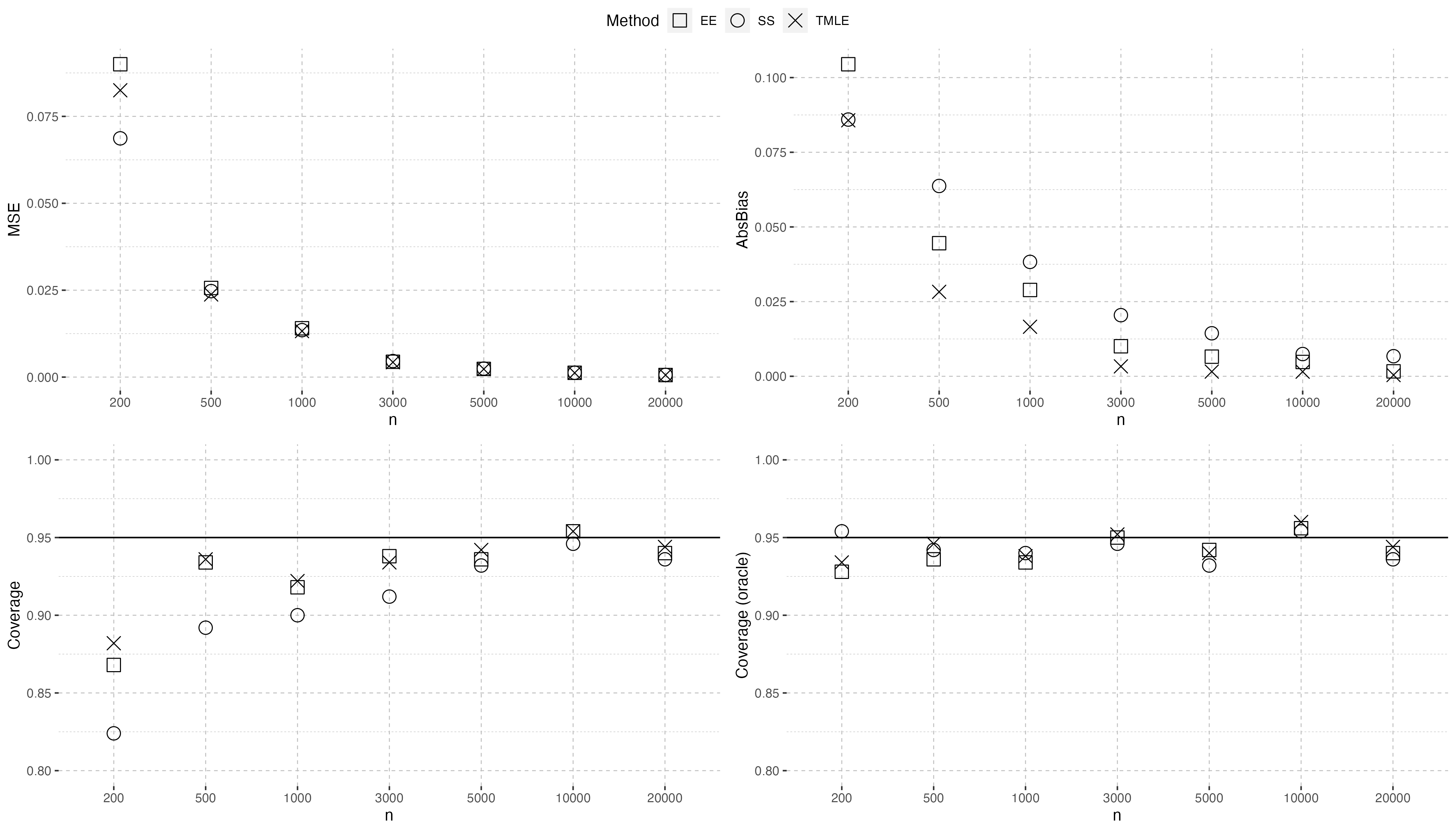}
    \caption{Main performance metrics of the estimators of the Unscaled VIM parameter }
    \label{p3}
\end{figure}

\noindent For the unscaled VIM (VIMa) parameter $\Psi_2(\tilde{Q}_0)$, Figure~\ref{p3} shows that, for all three estimators, the bias and MSE decrease and the coverage increases as sample size increases. TMLE outperforms EE and SS estimators in terms of lower bias and better coverage across almost all sample sizes. Three estimators have similar MSE and variance. Their oracle coverages are all close to $0.95$. 

\subsubsection{Results on the Scaled VIM Parameter} 

\begin{figure}[ht]
    \centering
    \includegraphics[scale = 0.35]{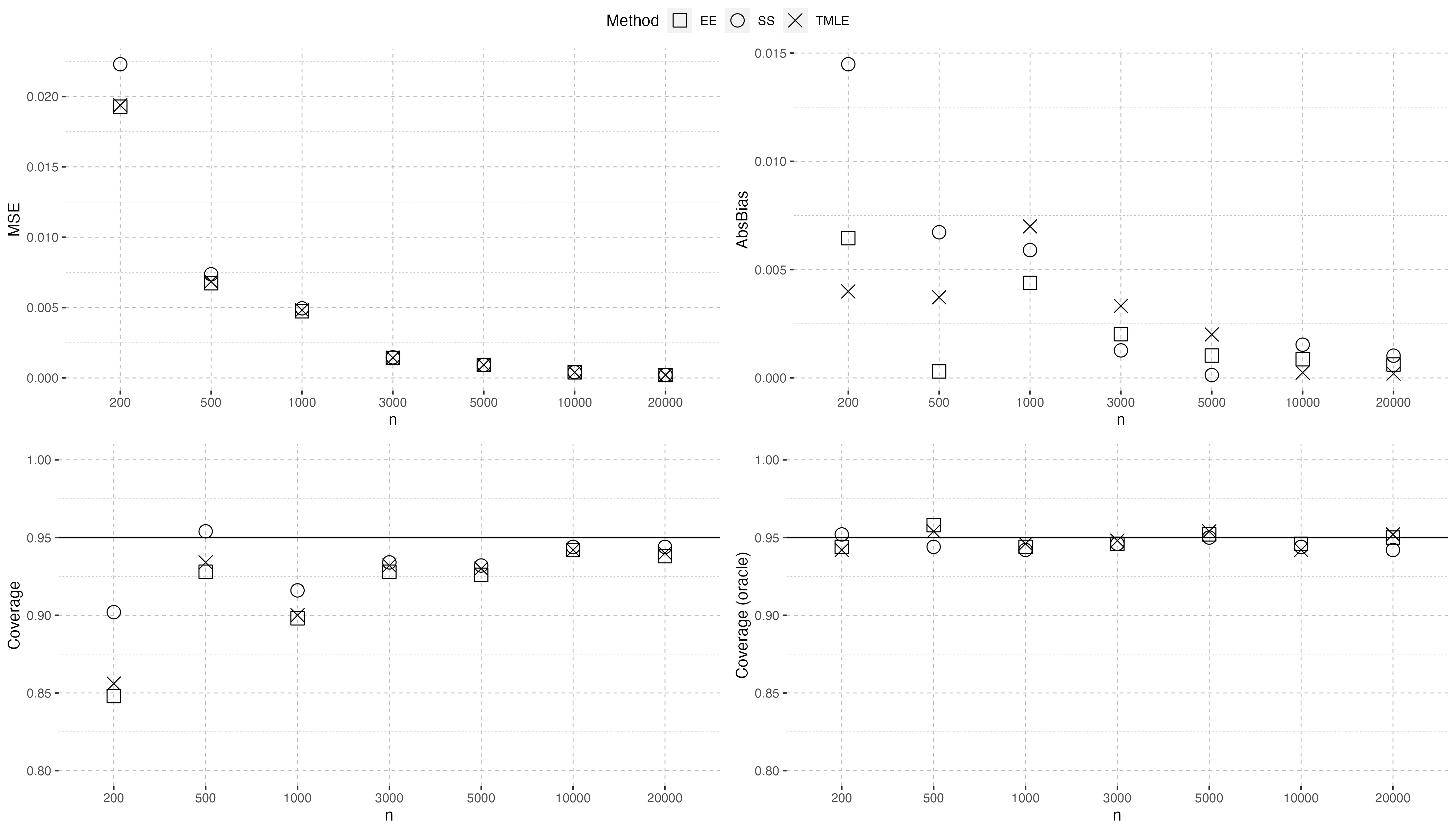}
    \caption{Main performance metrics of the estimators of the Scaled VIM parameter}
    \label{p5}
\end{figure}

\noindent For the scaled VIM (VIMb) parameter $\Psi_3(\breve{Q}_0)$, Figure~\ref{p5} shows that TMLE still performs the best asymptotically. When sample sizes are small/moderate, EE estimates sometimes have smaller bias. But the reduction in bias is negligible given the magnitude of the standard error. So the coverage of the CI estimator will still be dominated by the accuracy of standard error estimates, which is reflected in the coverage (TMLE still got better coverage although it has slightly larger bias than EE). Their oracle coverages are all close to $0.95$.

\section{Data Analysis}

In this section, we apply the EE and TMLE methods for VIMa to a data from the AIDS Clinical Trials Group Protocol 175 (ACTG175)~\citep{HammerTCNMO1996}. The data is available from the \texttt{speff2trial} R package. The are $2139$ patients infected with HIV in the original trial. They were randomized into four treatment groups: (i) zidovudine (ZDV) monotherapy, (ii) ZDV+didanosine (ddI), (iii) ZDV+zalcitabine, and (iv) ddI monotherapy. We compare treatment groups (iv) and (ii) as in~\cite{LuVSOTO2013, CuiEHTEA2023, HinesVIMHA2022}. We consider a continuous outcome $Y$ defined as the CD4 count at $20\pm 5$ weeks. As in~\cite{HinesVIMHA2022}, we include $12$ baseline covariates, $5$ continuous: age, weight, Karnofsky score, CD4 count, CD8 count; and $7$ binary: sex, homosexual activity (y/n), race (white/non-white), symptomatic status (symptomatic/asymptomatic), history of intravenous drug use (y/n), hemophilia (y/n), and antiretroviral history (experienced/naive). \\

\noindent We use SL to obtain estimates of relevant portions of the target parameter VIMa. Specifically, for outcome regression and CATE estimation, we construct a SL library with a linear model, a generalized additive model~\citep{HastieGAMA1986}, a gradient boosting model~\citep{ChenXSTBA2016}, a random forest model~\citep{WrightRFIRM2017}, a multivariate adaptive regression splines model~\citep{FriedmanMARSM1991} and a HAL model. For the propensity score estimation, since it is a randomized trial, we use a simplified SL library with only a linear model, a gradient boosting model and a multivariate adaptive regression splines model. As for the choice of metalearners, we use non-negative least squares~\citep{LingSLSPD1977} for continuous response variables and nonlinear optimization via augmented Lagrange~\citep{TseIALQ1997} for discrete response variables. We use 10-fold cross-validation scheme for all SL models. In addition, given the dimensions of the data, to obtain more robust estimates, we follow the sample splitting scheme proposed in~\cite{HinesVIMHA2022} to obtain cross-validated SL estimates. \\

\begin{figure}[ht]
    \centering
    \includegraphics[scale = 0.4]{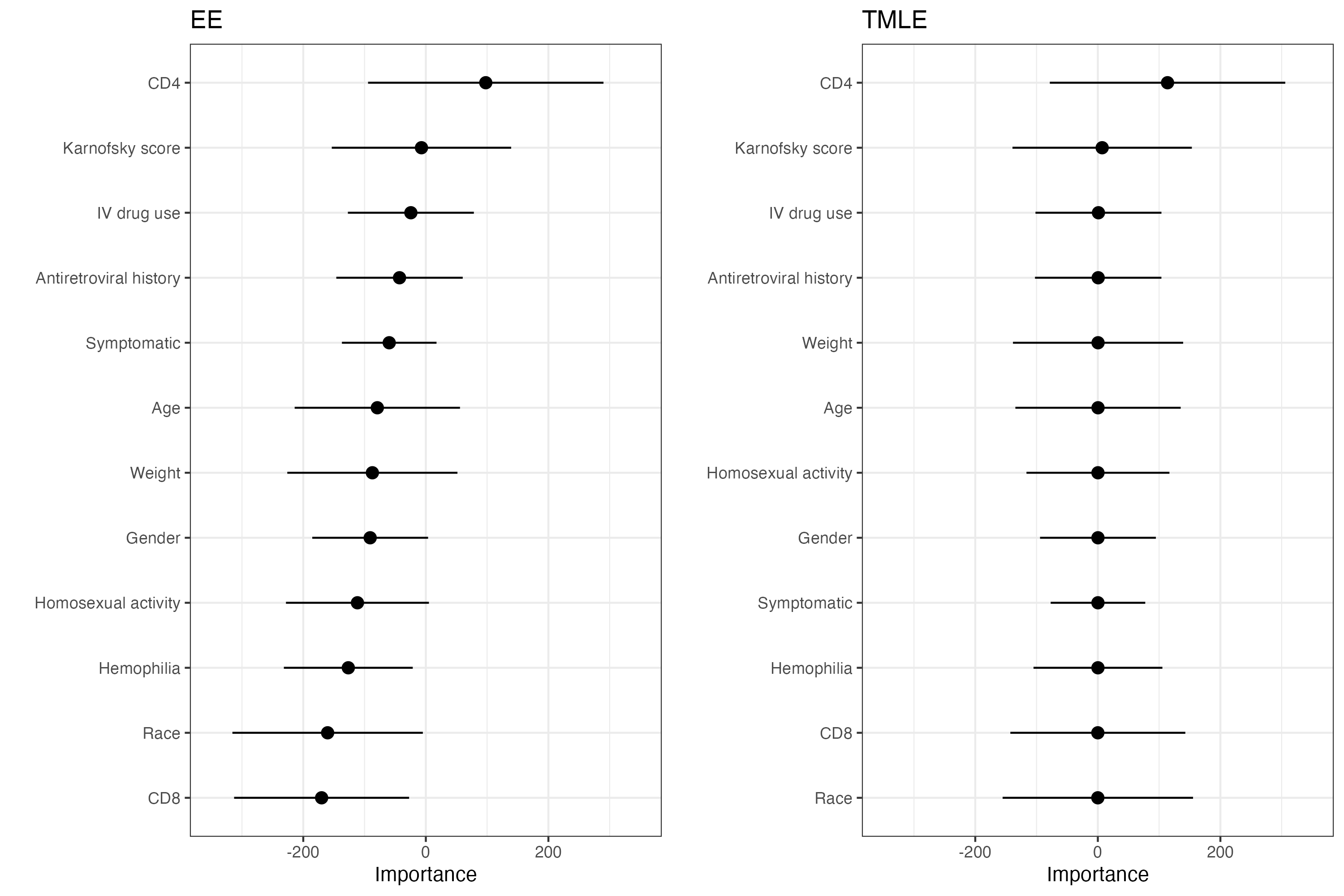}
    \caption{EE (left) and TMLE (right) VIMa results from the ACTG175 study with S-learner}
    \label{p7}
\end{figure}

\noindent Figure~\ref{p7} shows the VIMa estimates with EE and TMLE methods respectively. The ranking of variables using these two methods are similar. Both estimators agree that the CD4 count at the baseline is the most important covariate in the sense that it explains most treatment heterogeneity, beyond that already explained by the other covariates. This finding is consistent with the findings with VIMb parameter in~\cite{HinesVIMHA2022}. However, it remains inconclusive given the wide confidence intervals of both EE and TMLE for CD4 VIMa. Also, observe that the VIMa TMLE for all other covariates are very close to zero, which suggests potential homogeneity among these subgroups. A number of VIMa estimates from the EE method are negative , which is theoretically impossible as the true value of VIMa should always be non-negative given the definition. The second observation reflect that it is necessary to use a plug-in estimator such as TMLE to make sure the global constraint of the target parameter will always be respected. The analysis results with DR-learner reflect similar observations so we attached it in the Appendix.

\section{Conclusion}
In this article, we discuss three target parameters on heterogeneous treatment effect, one VTE parameter and two VIM parameters. We propose a new TMLE estimator for the unscaled VIM, and evaluate the performance of TMLE, SS and EE estimators through simulations. We find that TMLE has competitive performance in terms of lower bias and better coverage in most settings. The application on the AIDS trial data also suggest that TMLE, as a plug-in estimator, gives more robust results which always respect the global constraint.\\

\noindent In addition, we analyze the remainder term of the difference between the TMLE and the VIM parameter. The exact remainder does not have the doubly robust structure. As a result, any efficient estimator will not be doubly robust, and specifically, consistent estimation is not achieved with consistent estimation of treatment mechanism alone. Thus, we recommend to use robust and fast-convergent algorithms such as HAL (or SL with HAL) for TMLE initial estimation to ensure that we can better control the remainder term, thereby improve the finite sample performance.

\bibliography{vim}

\newpage
\appendix

\section{Theorem 1}
\noindent We claim that, under assumptions A1, A2 and A3, this TMLE satisfies:\\

\noindent \textit{Claim.}
\begin{align}
     &\Psi_2(\tilde{Q}_n^*) - \Psi_2(\tilde{Q}_0) = P_n D^*_{\Psi_2, \PP_0} + o_p(n^{-\frac{1}{2}})
\end{align}

\noindent An estimator, for which the difference between itself and the truth behaves like an empirical mean of a influence score plus a second order remainder, is also known as an asymptotically linear estimator. 

\begin{proof}
To show equation (10), we first consider a generic plug-in estimator $\psi_{2,n}$. By the von Mises expansion and the Riesz-representation theorem, it is known that $\psi_{2,n} - \Psi_2(\tilde{Q}_0) = - P_0 D_{\PP_n}^* + R(\PP_n, \PP_0)$, where $P_0 D_{\PP_n}^* = E(D_{\PP_n}^*)$ under $\PP_0$ and $R(\PP_n, \PP_0)$ is the remainder term~\cite{HinesDSLBJ2022}. Then with a few algebraic tricks, one can show that:
\begin{align}
\psi_{2,n} - \Psi_2(\tilde{Q}_0) &= - P_0 D_{\PP_n}^* + R(\PP_n, \PP_0)\\
&= (P_n - P_0)(D_{\PP_n}^*-D^*_{\Psi_2, \PP_0}) + P_n D^*_{\Psi_2, \PP_0} - P_n D_{\PP_n}^* + R(\PP_n, \PP_0)\\
&= P_n D^*_{\Psi_2, \PP_0} + (P_n - P_0)(D_{\PP_n}^*-D^*_{\Psi_2, \PP_0}) - P_n D_{\PP_n}^* + R(\PP_n, \PP_0)
\end{align}

\noindent Now, to obtain equation (10), we need to show that $(P_n - P_0)(D_{\PP_n}^*-D^*_{\Psi_2, \PP_0}) - P_n D_{\PP_n}^* + R(\PP_n, \PP_0)$ is $o_p(n^{-\frac{1}{2}})$. One sufficient condition is that each term is $o_p(n^{-\frac{1}{2}})$. \\

\noindent First, for the empirical process term $(P_n - P_0)(D_{\PP_n}^*-D^*_{\Psi_2, \PP_0})$, one needs two assumptions. \\

\noindent \textit{Assumption 1.} $L^2$ consistency: $||D_{\PP_n}^*-D^*_{\Psi_2, \PP_0}||_{L^2(\PP_0)} \overset{P}{\rightarrow} 0$.\\

\noindent \textit{Assumption 2.} Donsker condition: $D_{\PP_n}^*$ falls into a Donsker class of cadlag functions with finite sectional variation norm. \\

\noindent Alternatively, the second assumption can be replaced by using sample splitting. With \textit{Assumptions 1} and \textit{2}, one can immediately obtains $(P_n - P_0)(D_{\PP_n}^*-D^*_{\Psi_2, \PP_0}) = o_p(n^{-\frac{1}{2}})$ by Lemma 19.24 in ~\cite{VaartAS1998}.\\

\noindent Second, for the $P_n D_{\PP_n}^*$ term, we solve it through the TMLE update step. We assume that we have $P_n D_{\PP_n}^* = o_p(n^{-\frac{1}{2}})$ for now and have a more through discussion on this in the following paragraphs. \\

\noindent Finally, we want to show that $R(\PP_n, \PP_0)$ is also $o_p(n^{-\frac{1}{2}})$. Notice $\Psi_2(\tilde{Q}_0)$ can also be expressed as:
\begin{align}
    \Psi_2(\tilde{Q}_0) &= \mbox{var}(\tau(W)) - \mbox{var}(\tau_s(W))\\
    &= E(\tau^2(W)) - E^2(\tau(W)) - E(\tau_s^2(W)) + E^2(\tau_s(W)) \\
    &= E(\tau^2(W)) - E(\tau_s^2(W))\\
    &= E[\tau^2(W)) + E(\tau_s^2(W)) - 2E(\tau_s(W)E(\tau(W)|W_s)]\\
    &= E(\tau^2(W)) + E(\tau_s^2(W)) - 2E(\tau(W)\tau_s(W))\\
    &= E(\tau^2(W) + \tau_s^2(W) - 2\tau(W)\tau_s(W))\\
    &= E((\tau(W) - \tau_s(W))^2)
\end{align}
Then, the remainder term for $\Psi_2(\tilde{Q}_0)$ is:
\begin{align}
    R(\PP_n, \PP_0) ={}&\psi_{2,n}  - \Psi_2(\tilde{Q}_0) + P_0 D_{\PP_n}^*  \\\notag
    ={}&\psi_{2,n} - E((\tau(W) - \tau_s(W))^2) + \\
    &{} E[2(\tau(W) - \tau_s(W))\frac{2A-1}{g(A|W)}(Y - \bar{Q}(A, W)) + (\tau(W) - \tau_s(W))^2 - \psi_{2,n}] \\\notag
    ={}&- E((\tau(W) - \tau_s(W))^2) + \\
    &{} E[2(\tau_n(W) - \tau_{s,n}(W))\frac{2A-1}{g_n(A|W)}(Y - \bar{Q}_n(A, W)) + (\tau_n(W) - \tau_{s,n}(W))^2] \\\notag
    ={}&E((\tau_n(W) - \tau_{s,n}(W))^2 - (\tau(W) - \tau_s(W))^2) + \\
    &{} E\{E[2(\tau_n(W) - \tau_{s,n}(W))\frac{2A-1}{g_n(A|W)}(\bar{Q}(A, W) - \bar{Q}_n(A, W))|W]\} \\\notag
    ={}&E((\tau_n(W) - \tau_{s,n}(W))^2 - (\tau(W) - \tau_s(W))^2) + \\
    &{} E\{2(\tau_n(W) - \tau_{s,n}(W))[\frac{g(1|W)}{g_n(1|W)}(\bar{Q}(1, W) - \bar{Q}_n(1, W)) - \frac{g(0|W)}{g_n(0|W)}(\bar{Q}(0, W) - \bar{Q}_n(0, W))]\}\\\notag
    ={}&E((\tau_n(W) - \tau_{s,n}(W))^2 - (\tau(W) - \tau_s(W))^2) + \\\notag
    &{} E\{2(\tau_n(W) - \tau_{s,n}(W))[\frac{g(1|W)}{g_n(1|W)}(\bar{Q}(1, W) - \bar{Q}_n(1, W)) - \\
    &{} \frac{g(0|W)}{g_n(0|W)}(\bar{Q}(0, W) - \bar{Q}_n(0, W))]\}\\\notag
    ={}&E((\tau_n(W) - \tau_{s,n}(W))^2 - (\tau(W) - \tau_s(W))^2) + 2(\tau(W) - \tau_n(W))(\tau_n(W) - \tau_{s,n}(W))) + \\\notag
    &{} E\{2(\tau_n(W) - \tau_{s,n}(W))[\frac{g(1|W)-g_n(1|W)}{g_n(1|W)}(\bar{Q}(1, W) - \bar{Q}_n(1, W)) - \\
    &{}\frac{g(0|W)-g_n(0|W)}{g_n(0|W)}(\bar{Q}(0, W) - \bar{Q}_n(0, W))]\}\\\notag
    ={}&E((\tau(W) - \tau_{s,n}(W))^2 - (\tau(W) - \tau_s(W))^2 - (\tau(W) - \tau_n(W))^2(W)) + \\\notag
    &{} E\{2(\tau_n(W) - \tau_{s,n}(W))[\frac{g(1|W)-g_n(1|W)}{g_n(1|W)}(\bar{Q}(1, W) - \bar{Q}_n(1, W)) - \\
    &{}\frac{g(0|W)-g_n(0|W)}{g_n(0|W)}(\bar{Q}(0, W) - \bar{Q}_n(0, W))]\}\\\notag
    ={}&E((\tau_s(W) - \tau_{s,n}(W))^2- (\tau(W) - \tau_n(W))^2) + \\\notag
    &{} E\{2(\tau_n(W) - \tau_{s,n}(W))[\frac{g(1|W)-g_n(1|W)}{g_n(1|W)}(\bar{Q}(1, W) - \bar{Q}_n(1, W)) - \\
    &{}\frac{g(0|W)-g_n(0|W)}{g_n(0|W)}(\bar{Q}(0, W) - \bar{Q}_n(0, W))]\}
\end{align}

\noindent Observe that $R(\PP_n, \PP_0)$ can be decomposed into the two terms in equation (29). The second term is a standard ``double robust'' term which can be bounded a constant times $||\bar{Q}(A,W) - \bar{Q}_{n}(A,W)||_2||g(A|W) - g_n(A|W)||_2$. The details on bounding this term can be found in targeted learning literature on ATE and VTE~\cite{LaanTLCI2011, LevyTLEH2019}. Thus, in the scenario where one can always correctly specified one of the outcome regression model and the propensity score model (e.g. in a randomized controlled trial), this ``double robust'' remainder term can always converge faster than $o_p(n^{-\frac{1}{2}})$. Now, the challenging part is to control the first term $E((\tau_s(W) - \tau_{s,n}(W))^2- (\tau(W) - \tau_n(W))^2)$ since the integrand in the expectation is a difference of squares. To make sure this term is $o_p(n^{-\frac{1}{2}})$, we need that $\tau(W) - \tau_n(W)$ and $\tau_s(W) - \tau_{s,n}(W)$ are both $o_p(n^{-1/4})$ in $L^2(P)$ norm. This means, even in randomized controlled trial where the ``double robust'' remainder term can be eliminated, we still need good CATE estimates to control the first remainder term, thereby guarantee the performance of the estimator. In other words, this TMLE is not doubly robust as the TMLE for the ATE parameter. \\

\noindent With all three remainder terms controlled, we have that $R(\PP_n, \PP_0)$ is $o_p(n^{-\frac{1}{2}})$. Replace all $\PP_n$ with TMLE-updated $\PP_n^*$ in the equations above, we obtain equation (10) and complete the proof.\\
\end{proof}

\section{Decomposition of $D^*_{\Psi_2, \PP_0}$}

\noindent \textit{Claim.} 
$D^*_{W} = (\tau(W) - \tau_s(W))^2 - \Psi_2(\tilde{Q}_0)$.

\begin{proof}
Let $D^*_{\Psi_2, \PP_0} = D^*_{\PP_0,1} + D^*_{\PP_0,2}$, where $D^*_{\PP_0,1} = 2(\tau(W) - \tau_s(W))\frac{2A-1}{g(A|W)}(Y - \bar{Q}(A, W))$ and $D^*_{\PP_0,2} = (\tau(W) - \tau_s(W))^2 - \Psi_2(\tilde{Q}_0)$. Then
\begin{align}
    D^*_{W} &= \sqcap(D^*_{\Psi_2, \PP_0} | T_{W}) \\
    &= E(D^*_{\Psi_2, \PP_0}|W) \\
    &= E(D^*_{\PP_0,1}|W) + E(D^*_{\PP_0,2}|W) \\
    &= EE(D^*_{\PP_0,1}|A,W) + D^*_{\PP_0,2} \\
    &= 0 + D^*_{\PP_0,2}\\
    &= (\tau(W) - \tau_s(W))^2 - \Psi_2(\tilde{Q}_0)
\end{align}
\end{proof}

\noindent Then $D^*_{\Psi_2, \PP_0, Y|A,W}= D^*_{\Psi_2, \PP_0} - D^*_{W} = 2(\tau(W) - \tau_s(W))\frac{2A-1}{g(A|W)}(Y - \bar{Q}(A, W))$.\\

\noindent Similarly, we get $D^*_{\Psi_2, \PP_0, W_{s}|W_{-s}}$ by 
\begin{align}
    D^*_{\Psi_2, \PP_0, W_{s}|W_{-s}} &= \sqcap(D^*_W | T_{W_{s}|W_{-s}}) \\
    &= E(D^*_W|W) - E(D^*_W|W_{-s}) \\
    &= (\tau(W) - \tau_s(W))^2 -  E((\tau(W) - \tau_s(W))^2 | W_{-s})\\
    &= (\tau(W) - \tau_s(W))^2 - E(\tau^2(W) + \tau_s^2(W) - 2\tau(W)\tau_s(W) | W_{-s})\\
    &= \tau^2(W) + \tau_s^2(W) - 2\tau(W)\tau_s(W) - 
    E(\tau^2(W)|W_{-s}) - \tau_s^2(W) + 2\tau_s^2(W) \\
    &= \tau^2(W) - E(\tau^2(W)|W_{-s}) - 2\tau_s(W)(\tau(W) - \tau_s(W))
\end{align}

\noindent Finally, we have
\begin{align}
    D^*_{\Psi_2, \PP_0, W_{-s}} &= \sqcap(D^*_W | T_{W_{-s}}) \\
    &= E(D^*_W|W_{-s}) \\
    &= E((\tau(W) - \tau_s(W))^2 | W_{-s}) - \Psi_2(\tilde{Q}_0)\\
    &= E(\tau^2(W) + \tau_s^2(W) - 2\tau(W)\tau_s(W) | W_{-s}) - \Psi_2(\tilde{Q}_0)\\
    &= E(\tau^2(W)|W_{-s}) - \tau_s^2(W) - \Psi_2(\tilde{Q}_0)
\end{align}

\section{Other derivations}
\begin{align}
    ~~~~~~D^*_{\Psi_2, \PP_0} =~& (\varphi(O) - \tau_s(W))^2 - (\varphi(O) - \tau(W))^2 - \Psi_2(\tilde{Q}_0) \\
    =~& \varphi^2(O) + \tau_s^2(W) - 2\varphi(O)\tau_s(W) - \varphi^2(O) - \tau^2(W) + 2\varphi(O)\tau(W) - \Psi_2(\tilde{Q}_0) \\
    =~& 2\varphi(O) (\tau(W) - \tau_s(W)) + (\tau_s^2(W) - \tau^2(W)) - \Psi_2(\tilde{Q}_0) \\\notag
    =~& 2[\frac{2A-1}{g(A|W)}(Y - \bar{Q}(A, W)) + \bar{Q}(1, W) - \bar{Q}(0, W)](\tau(W) - \tau_s(W)) + \\
    &{} (\tau_s^2(W) - \tau^2(W)) - \Psi_2(\tilde{Q}_0) \\\notag
    =~& 2(\tau(W) - \tau_s(W))\frac{2A-1}{g(A|W)}(Y - \bar{Q}(A, W)) + \\
    &{} 2(\tau^2(W) - \tau(W)\tau_s(W)) + (\tau_s^2(W) - \tau^2(W)) - \Psi_2(\tilde{Q}_0)\\
    =~& 2(\tau(W) - \tau_s(W))\frac{2A-1}{g(A|W)}(Y - \bar{Q}(A, W)) + (\tau(W) - \tau_s(W))^2 - \Psi_2(\tilde{Q}_0)
\end{align}

\section{Results with double robust CATE estimation}

\begin{figure}[H]
    \centering
    \includegraphics[scale = 0.35]{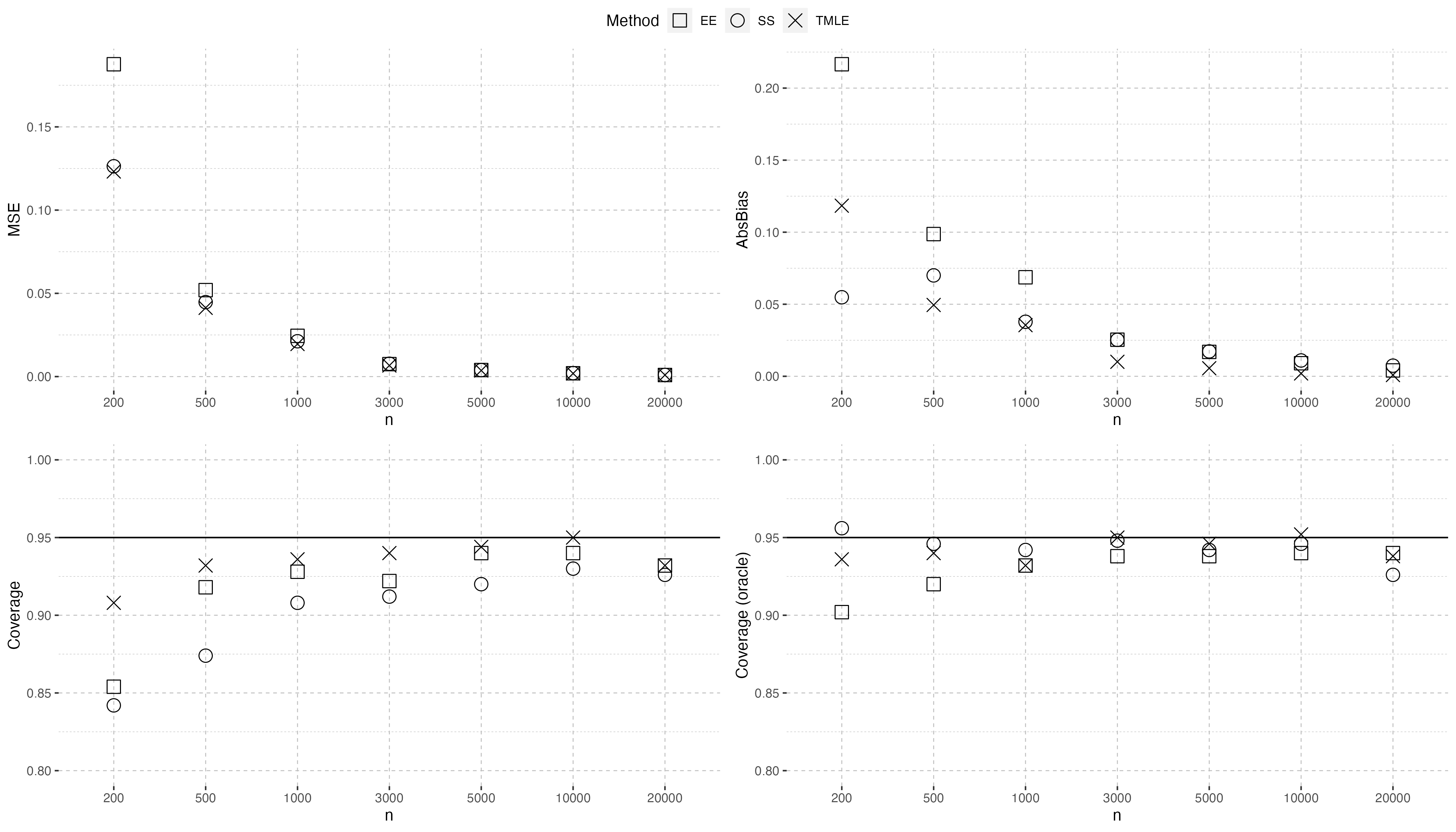}
    \caption{Main performance metrics of the estimators of the VTE parameter (DR-learner)}
    \label{p2}
\end{figure}

\begin{figure}[H]
    \centering
    \includegraphics[scale = 0.35]{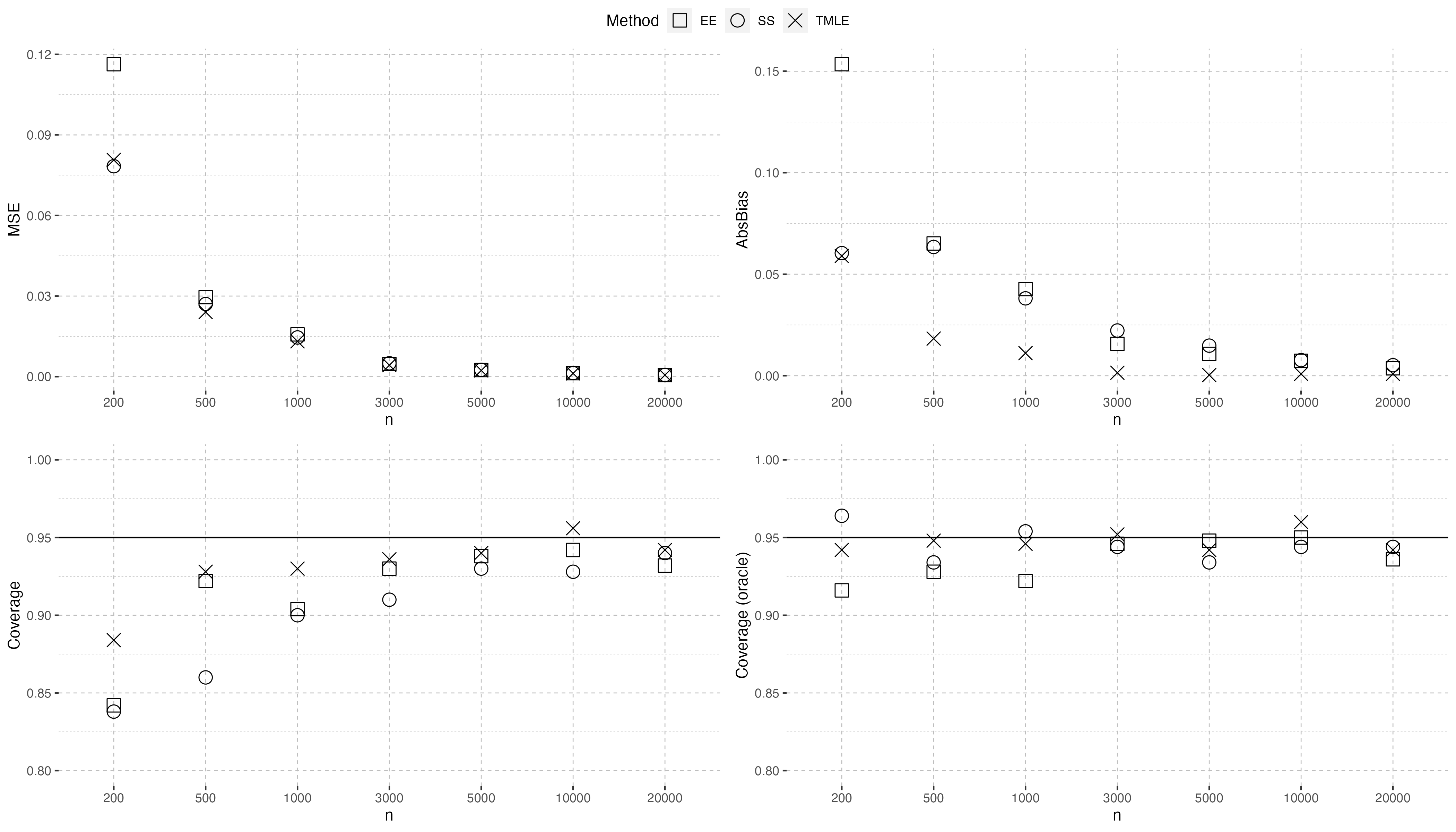}
    \caption{Main performance metrics of the estimators of the first VIM parameter (DR-learner)}
    \label{p4}
\end{figure}

\begin{figure}[H]
    \centering
    \includegraphics[scale = 0.35]{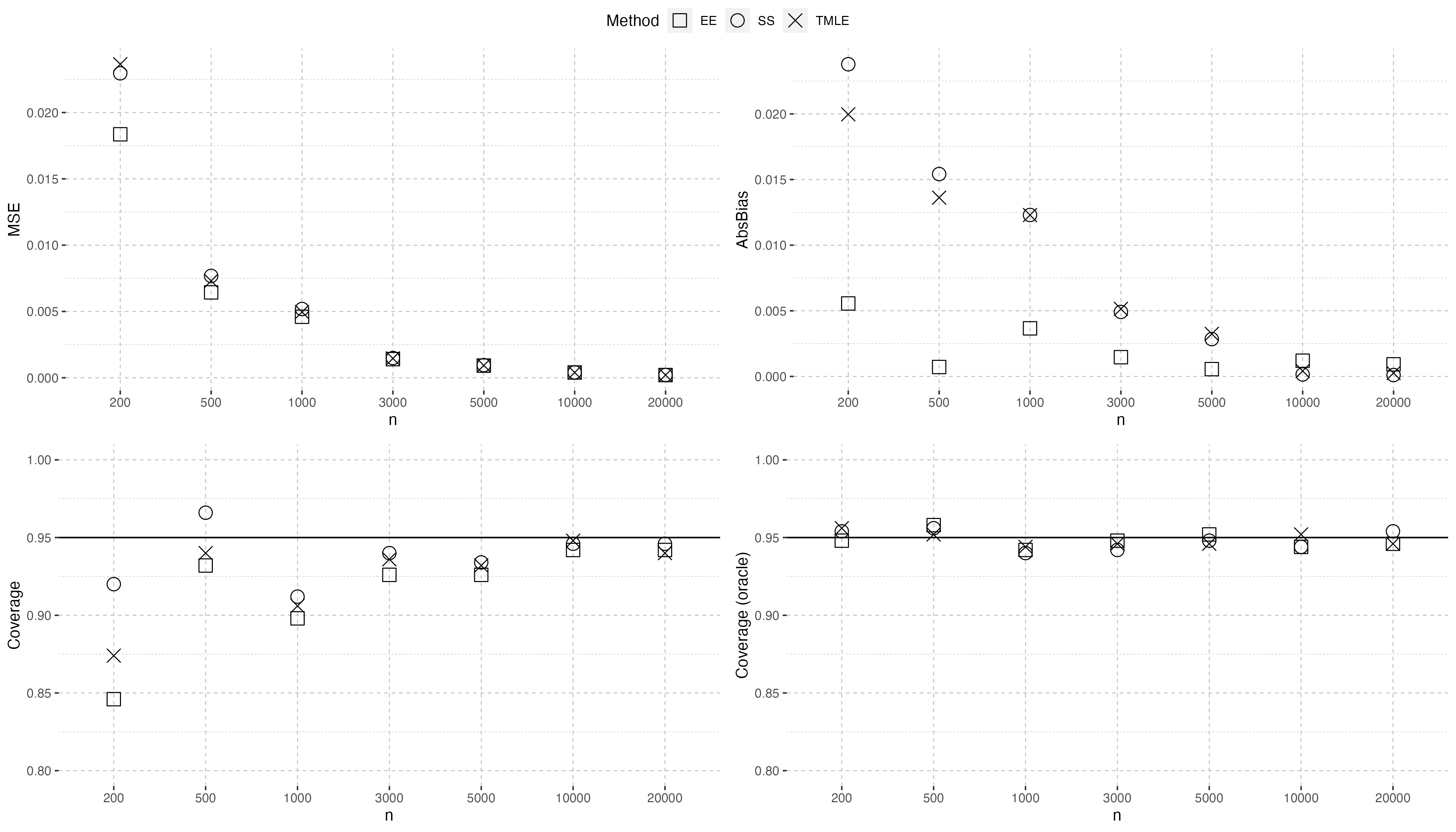}
    \caption{Main performance metrics of the estimators of the second VIM parameter (DR-learner)}
    \label{p6}
\end{figure}

\begin{figure}[H]
    \centering
    \includegraphics[scale = 0.4]{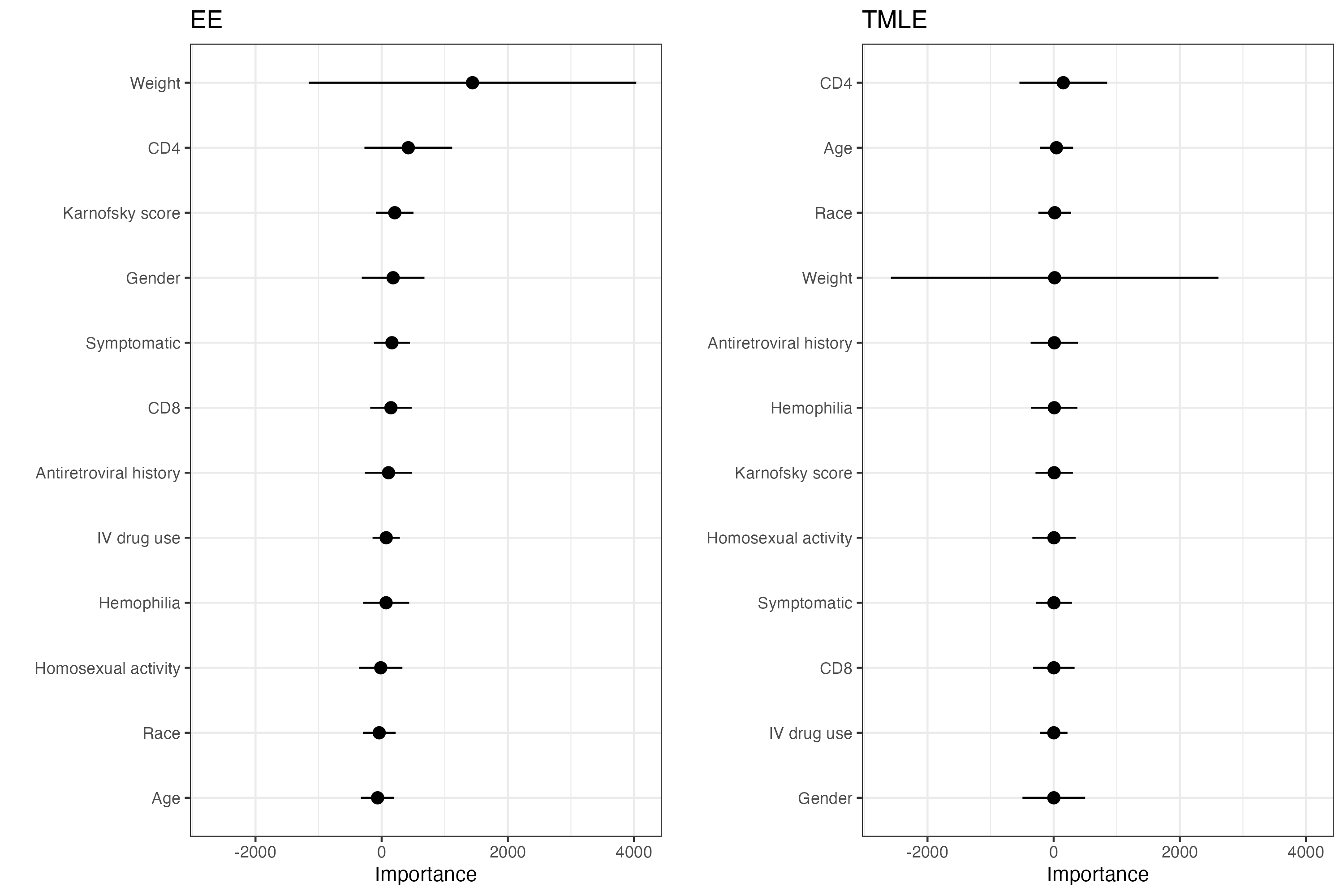}
    \caption{EE and TMLE VIMa results from the ACTG175 study with DR-learner}
    \label{p8}
\end{figure}

\end{document}